\documentclass[conference]{IEEEtran}
\IEEEoverridecommandlockouts
\usepackage{cite}
\usepackage{amsmath,amssymb,amsfonts}
\usepackage{algorithmic}
\usepackage{graphicx}
\usepackage{textcomp}
\usepackage{array}
\usepackage{mdwmath}
\usepackage{mdwtab}
\usepackage{eqparbox}
\usepackage{url}
\usepackage{supertabular}                                
\usepackage{caption}
\usepackage{subcaption}
\usepackage{float}
\usepackage{booktabs}
\usepackage{multirow}
\usepackage{graphicx}
\usepackage{color}
\usepackage{tabularray}
\usepackage{hhline}
\usepackage{wasysym}
\usepackage{rotating}
\usepackage{hyperref}
\usepackage{xcolor}
\usepackage{microtype}
\def\BibTeX{{\rm B\kern-.05em{\sc i\kern-.025em b}\kern-.08em
    T\kern-.1667em\lower.7ex\hbox{E}\kern-.125emX}}
\begin{document}

\title{Towards Threat Modelling of IoT Context-Sharing Platforms
\thanks{This work has been supported by the Cyber Security Research Centre Limited (CSCRC) whose activities are partially funded by the Australian Government’s Cooperative Research Centres Programme.}
}

\author{
   \IEEEauthorblockN{Mohammad Goudarzi\textsuperscript{\dag}\textsuperscript{\ddag}, Arash Shaghaghi\textsuperscript{\dag}\textsuperscript{\ddag}, Simon Finn\textsuperscript{\P}, Burkhard Stiller\textsuperscript{\S}, Sanjay Jha\textsuperscript{\dag}\textsuperscript{\ddag}}
\IEEEauthorblockA{\textsuperscript{\dag}School of Computer Science and Engineering, UNSW Sydney, Australia}
\IEEEauthorblockA{\textsuperscript{\ddag}Cyber Security Cooperative Research Centre (CSCRC)}
\IEEEauthorblockA{\textsuperscript{\S}Communication Systems Group, Department of Informatics, University of Zurich, Switzerland}
\IEEEauthorblockA{\textsuperscript{\P}Cisco}
}

\maketitle

\begin{abstract}
The Internet of Things (IoT) involves complex, interconnected systems and devices that depend on context-sharing platforms for interoperability and information exchange. These platforms are, therefore, critical components of real-world IoT deployments, making their security essential to ensure the resilience and reliability of these “systems of systems.” In this paper, we take the first steps toward systematically and comprehensively addressing the security of IoT context-sharing platforms. We propose a framework for threat modelling and security analysis of a generic IoT context-sharing solution, employing the MITRE ATT\&CK framework. Through an evaluation of various industry-funded projects and academic research, we identify significant security challenges in the design of IoT context-sharing platforms. Our threat modelling provides an in-depth analysis of the techniques and sub-techniques adversaries may use to exploit these systems, offering valuable insights for future research aimed at developing resilient solutions. Additionally, we have developed an open-source threat analysis tool that incorporates our detailed threat modelling, which can be used to evaluate and enhance the security of existing context-sharing platforms.

\end{abstract}


\section{Introduction}
\label{sec:introduction}
Context describes the state of an element, such as a person, place, application, or computing device, thus providing information on its characteristics and environment. Context information is often kept locally in IoT deployments and not shared, limiting the return on investment and utility of such deployments \cite{de2020context, ramachandran2019towards}. Consider the case of a smart city, where IoT devices and their data serve different application domains (e.g. traffic management, industries, etc). Each of these applications requires specific systems, and context-sharing enables such systems to understand context information across heterogeneous environments that may differ in format, data types, and specifications. In essence, context-sharing enables systems deployed in pervasive computational environments to have a common view and achieve interoperability.
\par
Context-sharing for IoT is an evolving domain. Large industry and government-funded projects have been working to design and build working platforms for context-sharing in IoT deployments. For instance, we have identified several projects under the EU Horizon 2020 program that aim to propose practical solutions for context-sharing platforms. An increasing number of academic research papers (see \cite{tiburski2015importance, de2020context, albouq2022survey} for surveys) have also been published aiming to address the challenges associated with developing practical and scalable solutions for interoperability and context-sharing in IoT deployments. \par 

As discussed by the authors in \cite{casadei2019development}, a generic IoT deployment can be classified into IoT Entities (which generate or utilise IoT Services), IoT Environments (the physical locations of IoT Entities), and IoT Services (the cyber-physical functions of IoT Entities). Context can then be regarded as the interdependencies among these components, encompassing explicit and implicit information about them. A context-sharing platform must facilitate complex application scenarios in IoT environments through a seamless and secure exchange of context information. The security of context information exchanges has unique characteristics and requirements. For instance, if an attacker obtains regular data from a communication channel, its meaning may not be easy to understand without the proper context. However, context information often represents a specific event semantically, making it much more understandable to an attacker. As discussed in previous research (e.g., \cite{de2020context}), and confirmed in our analysis (see \S\ref{qualitative_evaluation}), most context-sharing solutions do not have a systematic and holistic consideration of security risks. This is a significant gap and imposes a major challenge on the resiliency and reliability of large-scale IoT deployments -- i.e., while the interoperability of such a `system of systems' is essential, insecure deployments pose significant risks. 
\par

In this paper, we take the initial steps towards a comprehensive threat modelling of a generic context-sharing platform and propose a framework for it. Context-sharing platforms are developed using varying design architectures and are made up of different components and data exchange streams. In other words, IoT context-sharing platforms lack predefined standards and components \cite{de2020context}. Therefore, we first identified the key components and phases of a generic IoT context-sharing platform by analysing several well-known real-world projects and practical research papers. We then use the MITRE ATT\&CK to systematically analyse the full spectrum of threats that adversaries may use to target a generic context-sharing solution. The outcome of our analysis provides an elaborated threat model that can guide the security analysis of existing IoT context-sharing platforms and help improve the security of future solutions through a `secure-by-design' approach. For this, we have developed an open-source threat analysis solution that can be used to evaluate existing solutions and assist in developing resilient CSPs. Note that while we highlight all techniques and sub-techniques applicable to the generic solution presented, we are not aiming to conduct a complete threat modelling process. A standard threat modelling process requires the decomposition of a specific system by identifying the various components, data flows, and trust boundaries. This is then followed by threat identification (threat vectors and events) and attack modelling (mapping the sequence of attacks and describing tactics and techniques) \cite{guide2021threatmodelling}. However, the varying architectures and implementations of IoT context-sharing solutions limit such threat modellings to a specific solution and do not allow the provision of a holistic understanding of security threats applicable to these platforms. As a result, we have used the MITRE ATT\&CK for our threat modelling, which provides a detailed list of potential threats that are applicable to generic IoT context-sharing systems. 
\par

The main contributions of this paper are as follows:
\begin{itemize}
    \item We have identified the key components of a generic IoT context-sharing solution by analysing and abstracting research papers and real-world industry projects.
    \item We have applied MITRE ATT\&CK on the identified components of a generic context-sharing solution and have identified and then grouped relevant techniques and sub-techniques for each component.  
    \item We have analysed the existing IoT context-sharing solutions and have identified the most prominent industry-funded projects and relevant research papers. We then applied our proposed framework and evaluated the security of these solutions. We reported on our qualitative evaluation of related work and highlighted research gaps and security challenges to motivate future work.
    \item We have developed an open-source threat analysis solution that enables developers and researchers to evaluate context-sharing solutions using our proposed framework.
\end{itemize}

\section{Background}
\label{background}

\subsection{MITRE ATT\&CK}
\label{related_work:MITRE}
The MITRE Adversarial Tactics, Techniques, and Common Knowledge (ATT\&CK) repository documents real-world observations of adversarial tactics and methodologies for global access \cite{Mitre}. MITRE ATT\&CK is a model and knowledge base for understanding adversary behaviour, covering attack lifecycle stages and generic platforms used to target systems. The strategy consists of two main components: tactics, which represent an attacker's goals, and techniques, which demonstrate how to achieve them. MITRE ATT\&CK has three main domains: 1) enterprise, which addresses adversarial behaviour in enterprise systems like cloud, networks, OS, and containers; 2) mobile, which addresses aggressive behaviour in mobile environments; and 3) Industrial Control Systems (ICS). Additionally, the ATT\&CK matrix offers \textit{mitigations}, security concepts and tools that prevent adversaries from executing specific techniques or sub-techniques. Analysing these mitigations for different techniques helps identify attack patterns and choose security controls from security standards.
\subsection{Context-Sharing Solutions}
We have divided the related work into two categories. The `Academic Research' category involves papers addressing theoretical and practical requirements for context-sharing solutions. The `Industry Projects' category includes solutions that are funded by major industry and government schemes (e.g., EU Horizon 2020) with publicly accessible deliverables and resources that cover various practical aspects of a context-sharing solution for IoT deployments.  
\subsubsection{Academic Research}
We have identified practical research papers on context-sharing platforms within the IoT domain and have shortlisted ten research papers presenting frameworks and practical solutions. Moreover, we have studied whether these works contain security considerations, revealing that only two works have developed or considered security mechanisms or features. Table~\ref{tab:evaluation:securityconsideration-papers} presents an overview of these papers.
\begin{table*}[]
\centering
\caption{Qualitative Comparison of Research Papers}
\label{tab:evaluation:securityconsideration-papers}
\resizebox{\textwidth}{!}{%
\begin{tabular}{|c|c|c|c|c|c|c|c|c|c|c|}
\hline
References / Year & \cite{faieq2017c2iot} / (2017) & \cite{forkan2015bdcam} / (2017) & \cite{huru2018bigclue} / (2018) & \cite{fortino2018agent} / 2018 & \cite{hassani2018context} / 2018 & \cite{liu2019scents} / 2019 & \cite{zhang2020demand} / 2020 & \cite{pereira2022platform} / 2022 & \cite{borges2023taming} / 2023 & \cite{papadakis2023comdex} / 2023 \\ \hline
Domain & Smart City & Smart Healthcare & Generic & Smart City & Smart City & Smart City & Smart Healthcare & Smart City & Generic & Smart City \\ \hline
Security Consideration & No  & No  & No  & No  & Yes  & No  & No  & Yes  & No  &  No  \\ \hline
\end{tabular}%
}
\vspace{-0.25cm}
\end{table*}
\par
Faieq et al.~\cite{faieq2017c2iot} proposed a cloud-based framework to deliver IoT context-aware services within smart cities. It employs a three-layer infrastructure to facilitate context-aware service provisioning, including Software as a Service (SaaS) for context visualization, Platform as a Service (PaaS) for data management, and Infrastructure as a Service (IaaS) for data collection. Forkan et al.~\cite{forkan2015bdcam} used a context-sharing platform in medical systems. Their system has five core components, including a monitoring system, a context aggregator, and a provider that performs reasoning to derive high-level contextual insights, context-aware cloud components that offer data and service management functionalities, service providers that offer context-aware services, and context data visualisation that presents contextual information for end users. Huru et al.~\cite{huru2018bigclue} proposed a generic data processing platform that consists of five fundamental layers: processing, messaging, storage, service registry, and visualization. This platform can be integrated with pre-existing frameworks to provide a cross-domain data layer solution for IoT. Fortino et al.~\cite{fortino2018agent} proposed a multi-layer architecture and a multi-agent solution designed for context-aware solutions. Their architecture framework consists of three primary layers: the high-level smart object architecture serving as the architectural agent model, the agent middleware layer that supports various agent platforms for different devices, and a programming and management layer. This architecture accommodates multiple communication models, such as message passing and publish/subscribe, along with a rule-based context reasoning model. Liu et al.~\cite{liu2019scents} presented a framework designed for context-sharing among proximate devices, relying on their geographical location within smart city scenarios. This framework sits between the hardware and application layers, comprising two main components: the neighbourhood agent, responsible for continuously detecting nearby devices, and the collaboration agent, which handles the queries for contextual information, its processing, and subsequent distribution. Zhang et al.~\cite{zhang2020demand} introduced a middleware for on-demand application deployments in resource-rich edge and cloud environments. The raw contextual data is modelled based on ontology into two categories: the external environment context, such as time, and the internal system context, such as server status. These attributes are used to trigger adjustments in application deployments. Borges et al.~\cite{borges2023taming} introduced a middleware solution to simplify the interaction between application developers and an IoT platform, requiring minimal code. This framework employs proxies to represent virtual entities within the IoT platform. These proxies effectively manage the complexities associated with synchronous and publish/subscribe interaction patterns when interfacing with the IoT platform. They provide a convenient way to bypass the necessity of understanding the IoT platform's specific API and data model. This framework can seamlessly integrate with various context-sharing platforms (e.g., FIWARE). Papadakis et al.~\cite{papadakis2023comdex} presented a context-aware federated framework and IoT platform to facilitate efficient data exchange between IoT devices and applications. This framework consists of three essential components: the federation topology component, responsible for managing the connections between different brokers, the knowledge-base component, which corresponds to the information model, and the action handler component, which interacts with clients to facilitate various data exchange operations. This framework uses a publish/subscribe model for message exchange between communities through a hierarchical federated topology and employs an advertisement-based mechanism. Hassani et al.~\cite{hassani2018context} introduced a semantic-based context-sharing platform for the smart city, composed of four main components: a security and communication manager, a context query engine, a context storage management system, and a context reasoning engine. Pereira et al.~\cite{pereira2022platform} presented a distributed framework leveraging the FIWARE context-sharing platform \cite{cirillo2019standardFIWARE}. This framework is structured in two layers: a) the platform middleware and b) the infrastructure middleware.
\subsubsection{Industry Projects}
Table~\ref{tab:evaluation:securityconsideration-Industry} presents an overview of the projects reviewed. Among EU FP7 projects, three projects mainly focus on context-sharing solutions. The main objective of the Internet of Things Architecture (IoT-A)\footnote{https://cordis.europa.eu/project/id/257521} is to shape an architectural reference model for the interoperability of the IoT system, providing design principles and guidelines for protocols, interfaces, and algorithms. Furthermore, developing a data format for resource-constrained environments minimises network traffic and interaction frequency for more efficient interoperability. The main goal of the COllaboration and INteroperability for networked enterprises (COIN)\footnote{https://cordis.europa.eu/project/id/216256} project is enabling integration of existing and new enterprise interoperability and collaboration services. COIN offers a semantic-based web solution for interoperability, employing an ontology-based approach for semantic processing. Additionally, COIN features a knowledge-based system that stores information about various entities, including devices and resources. The main goal of the Interoperability of Data and procedures In large-scale multinational disaster Response Actions (IDIRA)\footnote{https://cordis.europa.eu/project/id/261726} is to facilitate information exchange and interoperability of services and to offer decision support to local and international parties for crisis management.
\par
Among HORIZON 2020 projects, six projects aimed at context-sharing. The primary mission of the Interoperability of Heterogeneous IoT Platforms (Inter-IoT)\footnote{https://cordis.europa.eu/project/id/687283} is to ensure interoperability among heterogeneous IoT platforms. It collaborates with other projects to achieve interoperability across various layers, including device, networking, middleware, application service, data and semantics, integrated IoT platform, and business levels. At the data and semantics level, INTER-IoT has created the Generic Ontology for IoT Platforms (GOIoTP) to facilitate semantic matching in IoT scenarios and streamline the process of context-sharing. Inter-IoT applied its framework to two application domains, including e-health and port transportation. The main goal of the Federated Interoperable Semantic IoT/cloud Testbeds and Applications (FIESTA)\footnote{https://cordis.europa.eu/project/id/643943} is to facilitate IoT experimentation by the interconnection and interoperability of various IoT systems, platforms, and testbeds. It employs a shared ontology, ensuring semantic consistency across various providers while using a standardized API, facilitating communication and granting access to information for IoT systems linked to it. The main goal of the Worldwide Interoperability for SEmantics IoT (Wise-IoT)\footnote{https://cordis.europa.eu/project/id/723156} is to provide interoperability concerning context information. It introduces a Global IoT Services (GIoTS) layer characterized by semantic interoperability to guarantee reliability and end-to-end security. Its Morphing Mediation Gateway (MMG) component plays a pivotal role, translating diverse protocols and data representations, while accommodating various ontologies. The main goal of the Smart End-to-end Massive IoT Interoperability, Connectivity and Security (SEMIoTICS)\footnote{https://cordis.europa.eu/project/id/780315} is to provide a pattern-driven solution for semantic interoperability within IoT environments, facilitating cross-layer adaptation for diverse smart objects across heterogeneous platforms. SEMIoTICS endeavours to address the critical challenge of ensuring secure and reliable actuation in IoT and industrial IoT applications by developing a pattern-driven framework. It encodes established dependencies among security, privacy, dependability, and interoperability, facilitating the seamless integration of smart objects. The Bridging the Interoperability Gap of the Internet of Things (Big IoT)\footnote{https://cordis.europa.eu/project/id/688038} introduces the BIG IoT API, a Web-based interface designed to enhance interoperability among IoT platforms. The Symbiosis of smart objects across IoT environments (SymbIoTe)\footnote{https://cordis.europa.eu/project/id/688156} introduces an IoT orchestration middleware that provides a unified perspective across diverse IoT platforms. It enables secure access to physical and virtualised IoT resources while facilitating hierarchical discovery and control across multiple platforms. Moreover, it fosters the federation of IoT controllers and resources, encouraging collaborative sensing and actuation tasks. The orchestration middleware builds upon existing standards for protocols and interfaces, incorporating proprietary platforms from industrial partners and open-source platforms like OpenIoT to ensure interoperability and compatibility across a wide range of IoT environments.
\begin{table}[t]
\centering
\caption{Qualitative Comparison of Industry Projects}
\label{tab:evaluation:securityconsideration-Industry}
\resizebox{\linewidth}{!}{%
\begin{tblr}{
  cells = {c},
  cell{2}{1} = {r=3}{},
  cell{5}{1} = {r=6}{},
  vlines,
  hline{1-2,5,11-13} = {-}{},
  hline{3-4,6-10} = {2-6}{},
}
Organization    & Project    & Main Domain                               & {Security\\Focus} & Budget (Euro) & Duration            \\
EU FP7          & IOT-A      & IoT                                       & Yes               & 18,321,198,00 & Sep 2010 - Nov 2013 \\
                & COIN       & Industry                                  & \_                & 16,118,535,00 & Jan 2008 - Dec 2011 \\
                & IDIRIA     & Crisis Management                         & Yes               & 10,836,072,86 & May 2011 - Apr 2015 \\
{HORIZON\\2020} & Inter-IOT  & IoT                                       & Yes               & 7,329,850,00  & Jan 2016 - Dec 2018 \\
                & FIESTA-IoT & {IoT - Environment\\~as a Service (EaaS)} & Yes               & 5,484,662,00  & Feb 2016 - Jun 2018 \\
                & Wise-IoT   & IoT                                       & Yes               & 1,776,810,75  & Jun 2016 - May 2018 \\
                & SEMIoTICS  & IoT                                       & Yes               & 4,995,915,00  & Jan 2018 - Dec 2020 \\
                & BIG IoT    & IoT                                       & Yes               & 7,999,882,50  & Jan 2016 - Dec 2018 \\
                & SymbIoTe   & IoT                                       & Yes               & 7,104,827,50  & Jan 2016 - Dec 2019 \\
FIWARE          & Orion      & IoT                                       & Yes               & \_            & Ongoing                  \\
ETSI            & ISG CIM    & Smart City                                & Yes               & \_            & Ongoing                  
\end{tblr}
}
\end{table}
\par
FIWARE offers a modular open-source framework designed to facilitate the development of IoT solutions. Serving as middleware within IoT environments, the FIWARE framework enables the seamless interconnection of devices and applications through various communication services. Among FIWARE's modules is the Orion Context Broker\footnote{https://fiware-orion.readthedocs.io/en/master/}, which plays a central role in managing context information. As a context broker, Orion receives context data from IoT environments and offers options to query or subscribe to specific contexts. It also features mechanisms for querying or subscribing to context based on geolocation, type, and format. As a context information repository, Orion contributes to achieving context interoperability within IoT ecosystems.
\par
The European Telecommunications Standards Institute (ETSI) has established a particular interest group dedicated to developing standards for context management systems. Known as the Industry Specification Group on Context Information Management (ISG CIM)\footnote{https://www.etsi.org/committee/1422-cim}, this group primarily concentrates on smart city applications. ISG CIM outlines a standard API to access context management systems tailored for smart city environments featuring heterogeneous data sources. 
\section{Methodology}
\label{Methodology}
Our framework is based on two critical elements for identifying potential threats in context-sharing platforms: 1) the main components and 2) the MITRE ATT\&CK matrix. 
\par
%
%
%
%
\begin{figure*}[!ht]
	\centering 
	\includegraphics[width=1\linewidth, height=5cm]{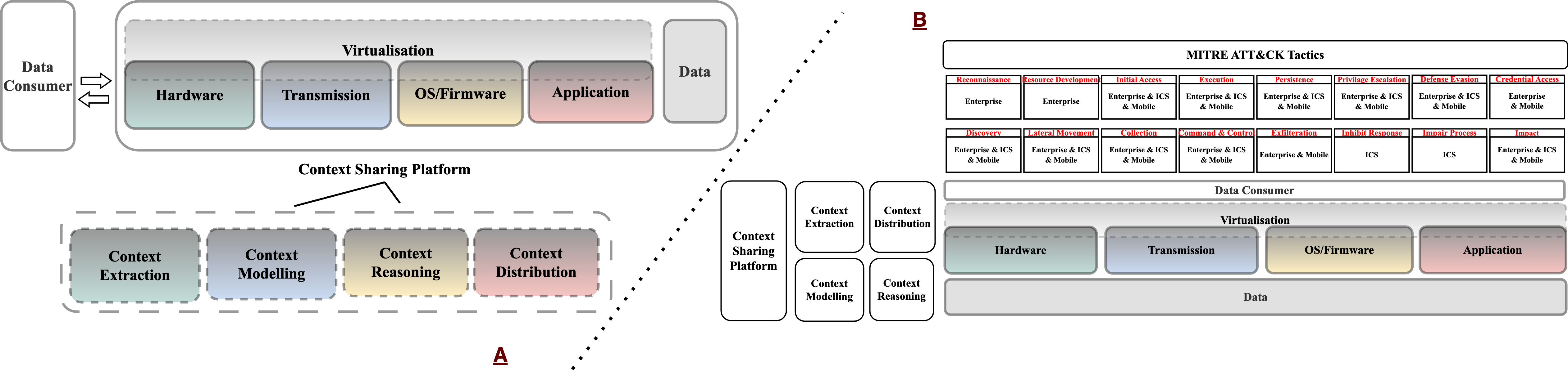}  
	\caption{A) The Main Components and Phases of a Generic Context-Sharing Platform, B) Visual Representation of Threat Modelling Framework.}
	\label{fig:mitre_context_sharing_methodology}
	\vspace{-.6cm}
\end{figure*}
\subsection{Context-Sharing Platform Components}
Context-sharing platforms offer a means to effectively share context information following four primary phases, including \textit{extraction, modelling, reasoning/inference, and distribution} \cite{perera2013context, de2020context}. In the extraction phase, the necessary raw data from various sources (e.g., user, infrastructure) within the designated environment is collected. Given the high diversity of sources and data types in the environment, it becomes essential to convert the extracted data into a standardized format, such as key-value pairs or ontologies. This conversion is the central objective of the modelling phase. In the reasoning phase, the information is transformed into an understandable context for end users or systems using different inference techniques, such as ontology and data fusion. Finally, the primary goal of the distribution phase is to disseminate the context information to the relevant parties.
\par
We have identified the primary components actively involved in context-sharing platforms, including \textit{data consumer, data, hardware, transmission, Operating System (OS)/firmware, application, and virtualisation}, as depicted in Fig. 1. We have identified these components by abstracting the key elements identified in real-world projects and research papers in context-sharing platforms, as described in \S\ref{background}.
\subsubsection{Data Consumer} refers to individuals within the designated environment (e.g., end user, administrator, staff) who either exist within the environment or engage with the context-sharing platform through actions (e.g., queries, utilizing dashboards).
\subsubsection{Hardware} refers to any physical devices and hardware components within context-sharing platforms, excluding network devices. Examples include servers, sensors, and physical components such as RAM and CPU.
\subsubsection{Transmission} pertains to physical network equipment, such as routers and switches, and all network protocols, including IP and routing. The rationale for distinguishing network equipment from hardware components lies in their distinct functions for transmission and communication, as networking devices serve a different purpose than other hardware elements. Additionally, a unique set of common threats is associated with networking equipment, which justifies its separation from other hardware components.
\subsubsection{OS/Firmware} includes any operating system, such as Mac, Linux, Windows, or firmware, which can be vendor-specific or open source and be operational on the physical hardware components.
\subsubsection{Application} It comprises any application, service, or process operating on OS, which may include examples such as word processing software and monitoring applications.
\subsubsection{virtualisation} It is a technology that spans hardware, networking, operating systems, and applications, offering advantages such as flexibility, efficient resource utilization, and isolation. It is a fundamental component in modern data centres and cloud computing environments. In the context of hardware, virtualisation allows the creation of Virtual Machines (VMs) on a single server or hardware platform, enabling these VMs to share the underlying hardware resources, such as CPU and RAM. Considering networking, virtualisation can extend to networking through technologies such as Virtual LANs (VLANs), virtual switches, and Software-Defined Networking (SDN). Virtual networks can be connected to VMs or containers, allowing isolated network environments and configurations independent of the physical network infrastructure. Regarding the OS, virtualisation enables the concurrent execution of various OS instances on the same physical hardware. This is possible through a Virtual Machine Monitor (VMM) or hypervisor, which abstracts the underlying hardware and provides a virtualised environment for each guest OS. Considering applications, virtualisation enables the deployment of applications within isolated virtualised environments. Each VM or container can run its applications and libraries, ensuring isolation and security.
\subsubsection{Data} 
\label{Methodology:data}
The data component contains raw information from the environment or contextual information accessible within the context-sharing platform. In addition, we include the medium utilized for data storage, such as a file, within this data category.  
\subsection{Mapping MITRE to Context-sharing Platform}
Regarding the primary phases of context-sharing platforms, context extraction and distribution predominantly involve interactions with the main components identified in the environment to extract contextual data or distribute the data among different components. These components include data consumers, hardware, transmission systems, OS/firmware, applications, and virtualisation. Conversely, the critical asset is the data within the context modelling and reasoning phases.
\par
To conduct a comprehensive threat modelling for each phase of context-sharing platforms, it is essential to determine the applicable domains within the MITRE ATT\&CK. To determine the relevant domains, we thoroughly examined all ATT\&CK domains, including enterprise, mobile, and ICS. Our investigation revealed that the primary components of context-sharing platforms span each of the MITRE ATT\&CK domains. Consequently, it becomes imperative to study all threats associated with these domains when conducting threat modelling for each phase of context-sharing platforms. Fig.~\ref{fig:mitre_context_sharing_methodology} illustrates how MITRE tactics (e.g., initial access, execution) are evaluated in relation to each phase of context-sharing platforms.

\begin{table*}[!ht]
\centering
\caption{Methodology: Data Consumer and Data Components}
\label{table:methodology:dataconsumer&data}
\resizebox{\linewidth}{!}{%
\begin{tblr}{
  row{1} = {c},
  row{2} = {c},
  row{19} = {c},
  column{2} = {c},
  cell{1}{1} = {c=2}{},
  cell{2}{1} = {c=2}{},
  cell{3}{1} = {r=16}{c},
  cell{19}{1} = {c=4}{},
  vlines,
  hline{1-3,19-20} = {-}{},
  hline{4-18} = {2-4}{},
}
{\textbf{Context-sharing}\\\textbf{Platform Phases}}          &                                           & \textbf{Context Extraction Context Distribution}                                                                                                                                                                                                                  & \textbf{Context Modelling Context Reasoning}                                                                                                                                                       \\
\textbf{Components}                                           &                                           & \textbf{Data Consumer}                                                                                                                                                                                                                                            & \textbf{Data}                                                                                                                                                                                      \\
\begin{sideways}\textbf{MITRE Tactics}\end{sideways}          & \textbf{Reconnaissance}                   & {T1589(001-003)(E),T1594(E),T1598(001-003(E),T1590(001)(E),\\T1591(001-004(E),T1597(001,002(E),T1593(001-003)(E),T1596(002-005)(E)}                                                                                                                               & Not Available                                                                                                                                                                                      \\
                                                              & {\textbf{Resource}\\\textbf{Development}} & T1586(001-003)(E),T1585 (001-003)(E)                                                                                                                                                                                                                              & Not Available                                                                                                                                                                                      \\
                                                              & {\textbf{Initial}\\\textbf{Access}}       & T1566(001-003)(E),T1078(001-004)(E),T0865(I)                                                                                                                                                                                                                      & T1190(E)                                                                                                                                                                                           \\
                                                              & \textbf{Execution}                        & Not Available                                                                                                                                                                                                                                                     & T1204(002)(E),T0863(I)                                                                                                                                                                             \\
                                                              & \textbf{Persistence}                      & T1098(001-005)(E),T1556(001-003,005-008)(E),T0859(I),T1078(001-004)(E)                                                                                                                                                                                            & Not Available                                                                                                                                                                                      \\
                                                              & {\textbf{Privilage}\\\textbf{Escalation}} & T1134(001-005)(E),T1078(001-004)(E)                                                                                                                                                                                                                               & Not Available                                                                                                                                                                                      \\
                                                              & {\textbf{Defense}\\\textbf{Evasion}}      & {T1134(001-005)(E),T1564(008)(E),T1070(001,002,008,009)(E),\\T1036(001,002,007)(E),T1556(001-003,005-008)(E),T1516(M)\\T1628(001-002)(M),T1629(002)(M),T1078(001-004)(E),T1550(001-004)(E)}                                                                       & {T1006(E),T1222(001,002)(E),T1564(001,004,005)(E),\\T1070(001-004)(E),T1036(006-008)(E),T1630(002)(M),~\\T1027(001,003,008,009,011)(E),T1406(001)(M),T0856 (I)}                                    \\
                                                              & {\textbf{Credential}\\\textbf{Access}}    & {T1557(001,003)(E),T1110(001-004)(E),T1551(001,002,004,005)(E),T1212(E),\\T1187(E),T1556 (001-003,005-008)(E),T1056(001-004)(E),T1111(E),T1621(E),\\T1040(E),T1003(001-006,008)(E),T1552(001,002,004,008)(E),T1517(M),\\T1414(M),T1634(001)(M),T1417(001-002)(M)} & T1552 (001,002) (E),T1414(M)                                                                                                                                                                       \\
                                                              & \textbf{Discovery}                        & {T1087(001-004)(E),T1217(E),T1040(E),T1201(E),T1069(001-003)(E),\\T1614(001)(E),T1033(E),T1430(001,002)(M),T0842(I)}                                                                                                                                              & T1217(E),T1619(E),T1083(E),T1120(E),T1420(M)                                                                                                                                                       \\
                                                              & {\textbf{Lateral}\\\textbf{Movement}}     & T1534(E),T1550(001-004)(E),T0859(I)                                                                                                                                                                                                                               & T1210 (E),~T1080,~T0867 (I)                                                                                                                                                                        \\
                                                              & \textbf{Collection}                       & {T1557(001,003)(E),T1056(001-004)(E),T1119(E),T1185(E),T1114(001-003),\\T1113(E),T1517(M),T1429(M,)T1616(M),T1414(M),T1417(001-002)(M),\\T1430(001,002)(M),T1636(001-004)(M),T1513(M),T0811(I)}                                                                   & {T1560(001-003)(E),T1119(E),T1115(E),T1005(E),T1025(E)\\T1602(001,002)(E),T1213(001-003)(E),T1039(E),T1530(E),\\T1074(001,002(E),T1532(M),T1414(M),T1533(M),T1409(M),\\T0802(I),T0811(I),T0893(I)} \\
                                                              & {\textbf{Command }\\\textbf{ Control}}    & T1001(001-003)(E),T1616(M)                                                                                                                                                                                                                                        & Not Available                                                                                                                                                                                      \\
                                                              & \textbf{Exfilteration}                    & Not Available                                                                                                                                                                                                                                                     & Not Available                                                                                                                                                                                      \\
                                                              & {\textbf{Inhibit}\\\textbf{Response}}     & T0892(I)                                                                                                                                                                                                                                                          & T0804(I),T0809(I),T0838(I),T0881(I)                                                                                                                                                                \\
                                                              & {\textbf{Impair}\\\textbf{Control}}       & Not Available                                                                                                                                                                                                                                                     & T0856(I)                                                                                                                                                                                           \\
                                                              & \textbf{Impact}                           & {T1531(E),T1486(E),T1565(001)(E),T1491(001,002)(E),T1640(M),T1616(M),\\T1471(M),T1643(M),T1516(M),T1582(M)}                                                                                                                                                       & {T1561(001,002)(E),T1490(E),T1489(E),T1485(E),T1486(E),\\T1565(001-003)(E),T1471(M),T1641(001)(M),T1582(M),\\T0826(I),T0832(I),T0882(I)}                                                           \\
E: Enterprise, M: Mobile, I: Industrial Control Systems (ICS) &                                           &                                                                                                                                                                                                                                                                   &                                                                                                                                                                                                    
\end{tblr}
}
\end{table*}

\begin{table*}
\centering
\caption{Methodology: Hardware, Transmission, OS/Firmware, and Application Components}
\label{table:methodology:hardware&transmission}
\resizebox{\linewidth}{!}{%
\begin{tblr}{
  row{1} = {c},
  row{2} = {c},
  row{19} = {c},
  column{2} = {c},
  cell{1}{1} = {c=2}{},
  cell{1}{3} = {c=4}{},
  cell{2}{1} = {c=2}{},
  cell{3}{1} = {r=16}{c},
  cell{19}{1} = {c=6}{},
  vlines,
  hline{1-3,19-20} = {-}{},
  hline{4-18} = {2-6}{},
}
                                                              &                                            & \textbf{Context Extraction $\&$ Context Distribution}                                                                                              &                                                                                                                                                                                                                                                                                                        &                                                                                                                                                                                                                                                                                                                                                                                                                                                                                                                                                                      &                                                                                                                                                                                                                                                                                                                                                                                                                       \\
                                                              &                                            & \textbf{Hardware}                                                                                                                                  & \textbf{Transmission}                                                                                                                                                                                                                                                                                  & \textbf{OS/Firmware}                                                                                                                                                                                                                                                                                                                                                                                                                                                                                                                                                 & \textbf{Application}                                                                                                                                                                                                                                                                                                                                                                                                  \\
\begin{sideways}\textbf{MITRE Tactics}\end{sideways}          & \textbf{Reconnaissance}                    & {T1595(001-003)(E),T1590(004)(E) \\T1592(001,004)(E),}                                                                                             & {T1595(001,002)(E),T1592(004)(E) \\T1590(001-006)(E),T1596(001,002)(E)}                                                                                                                                                                                                                                & T1592(003,004)(E)                                                                                                                                                                                                                                                                                                                                                                                                                                                                                                                                                    & {T1595(001-003)(E),T1584(001)(E)\\T1592(002-004)(E)}                                                                                                                                                                                                                                                                                                                                                                  \\
                                                              & {\textbf{Resource}\\\textbf{Development}}  & {T1650(E),T1583(003-005)(E) \\T1584(001,002,004,005)(E)\\T1588(005,006)(E),T1587(004)(E)}                                                          & {T1650(E),T1583(001)(E),T1587(004)(E) \\T1584(005)(E),T1588(005,006)(E)}                                                                                                                                                                                                                               & T1587(004)(E),T1588(005,006)(E)                                                                                                                                                                                                                                                                                                                                                                                                                                                                                                                                      & {T1583(002,008)(E),T1650(E) \\T1608(001-004)(E),T1588(001-006)(E) \\T1584(001)(E),T1587(001-004)(E)}                                                                                                                                                                                                                                                                                                                  \\
                                                              & {\textbf{Initial}\\\textbf{Access}}        & {T1566(001-003)(E),T1091(E)\\T1195(003)(E),T0862(I) \\T1474(002)(M),T0847(I)\\T0864(I),T0860(I),T0848(I)\\T1200(E),T1461(M),T1458(M)}              & {T1200(E),T1091(E),T1199(E),T0886(I)\\T0819(I),T0883(I),T0848(I),T0860(I)}                                                                                                                                                                                                                             & {T1091(E),T1195(003)(E),T1133(E) \\T1200(E),T1566(001-003(E) \\T1199(E),T1078(001)(E),T1461(M) \\T1458(M),T0819(I),T0886(I) \\T1474(002)(M),T0862(I)}                                                                                                                                                                                                                                                                                                                                                                                                                & {T1189(E),T1190(E),T1133(E),T0822(I) \\T1199(E),T1566(001-003)(E),T1200(E) \\T1195(001,002)(E),T0862(I),T0817(I) \\T1078(001-003)(E),T1456(M),T1461(M) \\T1458(M),T1474(001,003)(M),T0865(I) \\T0819(I),T0864(I)T0866(I),T0822(I),T0886(I)}                                                                                                                                                                           \\
                                                              & \textbf{Execution}                         & {T1651(E),T1106(E),T1072(E) \\T1603(M),T0858(I),T0807(I) \\T0871(I),T0834(I),T0974(I)}                                                             & {T1059(008)(E),T1106(E),T1072(E) \\T1204(001-003)(E),T1623(001)(M) \\T0858(I),T0807(I)}                                                                                                                                                                                                                & {T1059(001-009)(E),T1559 (001-003)(E),T0807(I) \\T1106(E),T1047(E),T0858(I),T1575(M),T1603(M) \\T1129(E),T1072(E),T1569(001,002)(E),T0807(I) \\T0834(I),T1623(001)(M),T1053(002,003,005,006)(E)}                                                                                                                                                                                                                                                                                                                                                                     & {T1651(E),T1059(001-009)(E),T1072(E)\\T1559(001-003(E),T1106(E),T1203(E)\\T1129(E),T1569(001,002)(E),T1610(E)\\T1053(002,003,005,006)(E)}                                                                                                                                                                                                                                                                             \\
                                                              & \textbf{Persistence}                       & {T1098(005)(E),T1078(001)(E)\\T1541(M),T0891(I),T0889(I)}                                                                                          & {T1136(001)(E),T1556(004)(E)\\T1205(001-002)(E),T0889(I)\\T1037(003)(E),T1078(001)(E)\\T1542(004-005)(E),T0891(I)}                                                                                                                                                                                     & {T1542(001-005)(E),T1546(001-016)(E)\\T1547(001-010,012-015)(E),T0857(I)\\T1037(001-005)(E),T1543(001-004)(E)\\T1574(001,002,004-013)(E) \\T1205(001-002)(E),T1078(001)(E)\\T1398(M),T1624(M),T1625(M) \\T1603(M),T0891(I),T0839(I)}                                                                                                                                                                                                                                                                                                                                 & {T1547(001-010,012-015)(E),T1197(E) \\T0873(I),T1037(001-005)(E),T1577(M)\\T1543(001-004)(E),T1176(E),T1133(E) \\T1137(001-006)(E),T1645(M),T1603(M)\\T1078(001-003)(E),T1546(001-016)(E) \\T1505(001-005)(E),T1053(002,003,005,006)(E)\\T1554(E),T0891(I),T1574(001,002,004-013)(E)}                                                                                                                                 \\
                                                              & {\textbf{Privilage }\\\textbf{Escalation}} & {T1055(001-009,011-015)(E)\\T1626(001)(M),T1078(001)(E) \\T1404(M),T1631(M),T0890(I) \\T0874(I)}                                                   & {T1037(003)(E),T1484(001-002)(E) \\T1055(001-009,011-015)(E) \\T1631(M),T0890(I)}                                                                                                                                                                                                                      & {T1547(001-010,012-015)(E),T1078(001)(E)\\T1543(001-004)(E),T1484(001-002)(E)\\T1611(E),T1548(001-004)(E),T0890(I)\\T1546(001-016)(E),T1068(E),T1631(M)\\T1404(M),T1574(001,002,004-013)(E)\\T1055(001-005,008,009,011-015)(E)\\T1053(002,003,005,006)(E),T1626(001)(M)}                                                                                                                                                                                                                                                                                             & {T1548(001-004)(E),T1404(M),T0874(I)\\T1037(001-005)(E),T0890(I),T1068(E)\\T1484(001-002)(E),T1543(001-004)(E)\\T1574(001,002,004-013)(E),T1078(001-003)(E)\\T1626(001)(M),T1631(M),T1546(001-016)(E)\\T1547(001-010,012-015)(E)}                                                                                                                                                                                     \\
                                                              & {\textbf{Defense}\\\textbf{Evasion}}       & {T1070(001,002,009)(E)\\T0858(I),T0851(I),T1407(M)\\T1631(M),T1078(001)(E)\\T1055(001-009,011-015)(E)\\T1014(E),T1562(006)(E)\\T1541(M),T1575(M),} & {T1610(E),T1484(001-002)(E)\\T1564(006)(E),T1599(001)(E)\\T1070(001-007)(E),T1407(M)\\T1604(M),T0858(I),T0851(I)\\T1562(003,004,006)(E)\\T1601(001,002)(E),T1631(M)\\T1542(004-005)(E),T1556(004)(E)\\T1550(001-003)(E),T1014(E)\\T1055(001-009,011-015)(E)\\T1205(001-002)(E)\\T1600(001-002)(E),}    & {T1548(001-004)(E),T1006(E),T1121(E),T1112(E)\\T1484(001-002)(E),T0858(I),T1216(E),T1014(E)\\T1222(001-002)(E),T1575(M),T1631(M)\\T1564(001-006,010)(E),T1647(E),T1202(E)\\T1497(001-003)(E),T0820(I),T0851(I),T1617(M)\\T1574(001,002,004-013)(E),T1620(E),T1221(E)\\T1562(001-003,006,009,010)(E),T1127(001)(E)\\T1070(001-006,008,009)(E),T0849(I)\\T1205(001-002)(E),T1218(001-005,007-014)(E)\\T1036(003,004,006,007)(E),T1629(003)(M)\\T1027(001-011)(E),T1550(001-003)(E)\\T1055(001-005,008,009,011-015)(E)\\T1078(001)(E),T1553(001-006)(E),T1632(001)(M),} & {T1548(001-004)(E),T1197(E),T0849(I)\\T1622(E),T1631(M),T1121(E),T1407(M)\\T1484(001-002)(E),T1216(E)\\T1564(007,009)(E),T1220(E)\\T1574 (001,002,004-013)(E)\\T1562(001,009)(E),T1620(E)\\T1070(001-006,008,009)(E),T1027(001-011)(E)\\T1553(001-006)(E),T1014(E)\\T1221(E),T1127(001)(E),T1629(001)(M)\\T1550(001-004)(E),T0851(I)\\T1218(001-005,007-014)(E)\\T1036(001)(E),T1617(M)\\T1078(001-003)(E),T0820(I),} \\
                                                              & {\textbf{Credential}\\\textbf{Access}}     & {T1110(001-004)(E)\\T1056(001)(E),T1111(E)\\T1552(001,004)(E)\\T1040(E),T1621(E)\\T1551(004)(E)}                                                   & {T1557(002,003)(E),T1556(004)(E)\\T1110(001-004)(E),T1111(E)\\T1551(004)(E),T1552(001,004)(E)\\T1056(001,002)(E),T1634(001)(M)\\T1621(E),T1040(E)}                                                                                                                                                     & {T1557(001)(E),T1110(001-004)(E)\\T1551(001,002,004)(E),T1212(E)\\T1056(001,002)(E),T1621(E),T1040(E)\\T1003(001-006)(E),T1649(E)\\T1558(001-004)(E),T1552(001-004)(E)\\T1634(001)(M),T1417(001-002)(M)}                                                                                                                                                                                                                                                                                                                                                             & {T1110(001-004)(E),T1212(E),T1634(001)(M)\\T1551(001-004)(E),T1111(E),T1621(E)\\T1414(M),T1056(001-004)(E),T1539(E)\\T1040(E),T1417(001-002)(M),T1649(E)\\T1003(004,007)(E),T1528(E),T1517(M)\\T1635(001)(M),T1552(001-004)(E)}                                                                                                                                                                                       \\
                                                              & \textbf{Discovery}                         & {T1652(E),T1040(E),T1120(E)\\T1426(M),T1614(001)(E)\\T1082(E),T0888(I),T0887(I)}                                                                   & {T1217(E),T1083(E),T1046(E)\\T1135(E),T1040(E),T1201(E),T0846(I)\\T1057(E),T1012(E),T1018(E)\\T1082(E),T1614(001)(E),T0887(I)\\T1016(001)(E),T1089(E),T1124(E)\\T1049(E),T1426(M),T1422(M)\\T1421(M),T0840(I),T0842(I),T0888(I)}                                                                       & {T1010(E),T1083(E),T1135(E),T1040(E)\\T1201(E),T1057(E),T1012(E),T1018(E)\\T1082(E),T1007(E),T1124(E),T1426(M)\\T1497(001-003)(E),T1420(M),T1424(M)\\T0888(I),T0888(I)}                                                                                                                                                                                                                                                                                                                                                                                              & {T1010(E),T1217(E),T1622(E),T0888(I)\\T1652(E),T1046(E),T1040(E),T1424(M)\\T1057(E),T1012(E),T1423(M),T1007(E)\\T1082(E),T0842(I),T1418(001)(M)\\T1518(001)(E)}                                                                                                                                                                                                                                                       \\
                                                              & {\textbf{Lateral}\\\textbf{Movement}}      & {T1210(E),T1091(E),T1072(E)\\T0891(I),T0812(I)}                                                                                                    & {T1210(E),T1091(E),T1072(E)\\T0886(I),T1550(001-003)(E)\\T0891(I),T0812(I)}                                                                                                                                                                                                                            & {T1210(E),T1570(E),T1563(001,002)(E)\\T1091(E),T1072(E),T1550(001-003)(E)\\T1080(E),T1021(001-006)(E),T1428(M)\\T1458(M),T0891(I),T0812(I),T0867(I),T0886(I)}                                                                                                                                                                                                                                                                                                                                                                                                        & {T1210(E),T1563(001,002)(E),T0812(I)\\T1072(E),T0866(I),T1080(E),T0843(I)\\T1550(001-004)(E),T1428(M),T0867(I)\\T1458(M),T0891(I),T0886(I)}                                                                                                                                                                                                                                                                           \\
                                                              & \textbf{Collection}                        & {T1056(001)(E),T1123(E)\\T1119(E),T1602(001,002)(E)\\T1213(001-003)(E),T1005(E)\\T1025(E),T1125(E),T1533(M)\\T1512(M),T08202(I),T0811(I)}          & {T1557(002,003)(E),T0852(I)\\T1119(E),T1602(001,002)(E)\\T1039(E),T1213(001-003)(E)\\T1005(E),T1638(M),T0830(I)\\T0811(I),T0893(I)\\T0887(I),T1056(001,002)(E),}                                                                                                                                       & {T1557(001)(E),T1056(001,002)(E)\\T1185(E),T1602(001,002)(E),T1005(E)\\T1213(001-003)(E),T1113(E),T1533(M)\\T1417(001-002)(M),T1513(M),T0802(I)\\T0811(I),T0893(I),T0868(I),T0877(I)\\T0801(I),T0852(I),T1119(E)}                                                                                                                                                                                                                                                                                                                                                    & {T1056(001-004)(E),T1123(E),T0802(I)\\T1119(E),T1185(E),T0845(I),T0861(I),T0877(I)\\T1602(001,002)(E),T1213(001-003)(E)\\T1005(E),T1113(E),T1517(M),T0801(I)\\T1414(M),T1533(M),T1513(M),T0893(I)\\T1417(001-002)(M),T1409(M),T0811(I)}                                                                                                                                                                               \\
                                                              & {\textbf{Command }\\\textbf{ Control}}     & {T1132(001,002)(E)\\T1001(001-003)(E)\\T1568(001-003)(E)\\T1573(001,002)(E)}                                                                       & {T1205(001-002)(E),T0884(I),T1509(M)\\T1071(001-004)(E),T1008(E),T1644(M)\\T1132 (001,002)(E),T1105(E),T1572(E)\\T1001(001-003)(E),T1104(E),T1571(E)\\T1568(001-003)(E),T1090(E),T1095(E)\\T1573(001,002)(E),T1521(001,002)(M)\\T1437(001)(M),T1481(001-003)(M)\\T0869(I),T1102(001-003)(E),T0885(I),} & {T1205(001-002)(E),T1092(E)\\T1132(001,002)(E),T1105(E)\\T1001(001-003)(E),T1568(001-003)(E)\\T1573(001,002)(E),T1104(E),T1095(E)\\T1521(001,002)(M),T1219(E)}                                                                                                                                                                                                                                                                                                                                                                                                       & {T1092(E),T1132(001,002)(E)\\T1001(001-003)(E)\\T1637(001)(M)\\T1573(001,002)(E),T1105(E)\\T1104(E),T1095(E),T1571(E)\\T1568(001-003)(E),T1219(E)}                                                                                                                                                                                                                                                                    \\
                                                              & \textbf{Exfilteration}                     & Not Available                                                                                                                                      & {T1020(001)(E),T1030(E),T1041(E)\\T1639(M),T1537(E),T1567(001-003)(E)\\T1048(001-003)(E),T1052(001)(E)\\T1011(001)(E),T1029(E),T1646(M)}                                                                                                                                                               & Not Available                                                                                                                                                                                                                                                                                                                                                                                                                                                                                                                                                        & T1537(E),T1052(001)(E)                                                                                                                                                                                                                                                                                                                                                                                                \\
                                                              & {\textbf{Inhibit}\\\textbf{Response}}      & {T0804(I),T0805(I),T0892(I)\\T0814(I),T0835(I),T0838(I)}                                                                                           & {T0804(I),T0805(I),T0892(I),T0851(I)\\T0814(I)}                                                                                                                                                                                                                                                        & {T0878(I),T0804(I),T0892(I),T0809(I),T0800(I),T0857(I)\\T0814(I),T0816(I),T0835(I),T0838(I),T0851(I),T0881(I)}                                                                                                                                                                                                                                                                                                                                                                                                                                                       & {T0878(I),T0804(I),T0892(I),T0881(I)\\T0814(I),T0816(I),T0838(I),T0851(I)}                                                                                                                                                                                                                                                                                                                                            \\
                                                              & {\textbf{Impair}\\\textbf{Control}}        & T0806(I),T0836(I)                                                                                                                                  & Not Available                                                                                                                                                                                                                                                                                          & T0839(I),T0855(I)                                                                                                                                                                                                                                                                                                                                                                                                                                                                                                                                                    & Not Available                                                                                                                                                                                                                                                                                                                                                                                                         \\
                                                              & \textbf{Impact}                            & {T1561(001,002)(E),T1531(E)\\T1495(E),T1496(E),T1464(M)\\T0917(I),T0826(I),T0827(I)\\T0828(I),T1485(E),T0879(I)}                                   & {T1531(E),T1485(E),T1486(E),T1464(M)\\T1565(001,002)(E),T1529(E),T0826(I)\\T1561(001,002)(E),T1489(E),T0817(I)\\T1499(002-004)(E),T1490(E),T0879(I)\\T1498(001,002)(E),1496(E),T0815(I)\\TT1641(001)(M),}                                                                                              & {T1486(E),T1565(001,003)(E),T0828(I),T0837(I)\\T1499(001,004)(E),T1490(E),T0880(I),T1495\\T1489(E),T1529(E),T1471(M),T0826(I),T0827(I),}                                                                                                                                                                                                                                                                                                                                                                                                                             & {T1485(E),T1486(E),T1496(E),T1464(M)\\T1565(001,003)(E),T1490(E),T1582(M)\\T1499(002-004)(E),T1643(M),T1642(M)\\T1498(001,002)(E),T1489(E),T0828(I)\\T0826(I),T1471(M)}                                                                                                                                                                                                                                               \\
E: Enterprise, M: Mobile, I: Industrial Control Systems (ICS) &                                            &                                                                                                                                                    &                                                                                                                                                                                                                                                                                                        &                                                                                                                                                                                                                                                                                                                                                                                                                                                                                                                                                                      &                                                                                                                                                                                                                                                                                                                                                                                                                       
\end{tblr}
}
\end{table*}

\begin{table*}
\centering
\caption{Methodology: Virtualisation Component}
\label{table:methodology:virtualisation}
\resizebox{\linewidth}{!}{%
\begin{tblr}{
  row{1} = {c},
  row{2} = {c},
  row{19} = {c},
  column{2} = {c},
  cell{1}{1} = {c=2}{},
  cell{2}{1} = {c=2}{},
  cell{3}{1} = {r=16}{c},
  cell{19}{1} = {c=3}{},
  vlines,
  hline{1-3,19-20} = {-}{},
  hline{4-18} = {2-3}{},
}
{\textbf{Context-sharing}\\\textbf{Platform Phases}}          &                                           & \textbf{Context Extraction Context Distribution}                                                                                                                                      \\
\textbf{Components}                                           &                                           & \textbf{Virtualisation}                                                                                                                                                               \\
\begin{sideways}\textbf{MITRE Tactics}\end{sideways}          & \textbf{Reconnaissance}                   & T1595(001-003)(E),T1592(004)(E),T1590(002)(E)                                                                                                                                         \\
                                                              & {\textbf{Resource}\\\textbf{Development}} & {T1583(003,005-007)(E),T1586(003)(E),T1584(003,006,007)(E),T1587(004)(E),T1585(003)(E),T1588(001,005,006)(E),T1608(001)(E),\\T1584(003)(E),T1587(004)(E),T1588(006)(E),T1608(001)(E)} \\
                                                              & {\textbf{Initial}\\\textbf{Access}}       & T1189(E),T1189(E),T1199(E),T1078(001-004)(E),T1190(E),T1133(E),T0822(I)                                                                                                               \\
                                                              & \textbf{Execution}                        & T1651(E),T1059(009)(E),T1648(E),T1204(001-003)(E),T1609(E),T1610(E),T1053(007)(E)                                                                                                     \\
                                                              & \textbf{Persistence}                      & T1098(001,003,004)(E),T1136(001-003)(E),T1078(001-004)(E),T1133(E),T1525(E),T1053(007)                                                                                                \\
                                                              & {\textbf{Privilage}\\\textbf{Escalation}} & {T1078(001-004)(E),T1611(E),T1068(E),\\T1053(007)(E)}                                                                                                                                 \\
                                                              & {\textbf{Defense}\\\textbf{Evasion}}      & {T1562(001-004,006-008)(E),T1070(001-004,006,009)(E),T1578(001-004)(E),T1078(001-004)(E),T1497(001-003)(E),T1535(E),T1612(E),\\T1610(E),T1550 (001-004)(E)}                           \\
                                                              & {\textbf{Credential}\\\textbf{Access}}    & {T1110(001-004)(E),T1551(001,002)(E),T1212(E),T1606(001-002)(E),T1056(001-003)(E),T1621(E),T1040(E),T1528(E),T1649(E),T1539(E),\\T1552(001,003-005,007)(E)}                           \\
                                                              & \textbf{Discovery}                        & {T1087(004)(E),T1580(E),T1538(E),T1526(E),T1619(E),T1046(E),T1040(E),T1201(E),T1069(003)(E),T1518(001)(E),T1082(E),T1049(E),\\T1124(E),T1497(001-003)(E),T1613(E)}                    \\
                                                              & {\textbf{Lateral}\\\textbf{Movement}}     & T1210(E),T1563(001,002)(E),T15550(001-004)(E),T1021(007)(E)                                                                                                                           \\
                                                              & \textbf{Collection}                       & T1056(001-003)(E),T1119(E),T1530(E),T1602(001,002)(E),T1213(001-003)(E),T1113(E)                                                                                                      \\
                                                              & {\textbf{Command }\\\textbf{ Control}}    & T1132(001,002)(E),T1001(001-003)(E),T1573(001,002)(E),T1219(E)                                                                                                                        \\
                                                              & \textbf{Exfilteration}                    & T1020(E)                                                                                                                                                                              \\
                                                              & {\textbf{Inhibit}\\\textbf{Response}}     & Not Available                                                                                                                                                                         \\
                                                              & {\textbf{Impair}\\\textbf{Control}}       & Not Available                                                                                                                                                                         \\
                                                              & \textbf{Impact}                           & T1531(E),T1485(E),T1486(E),T1565(001)(E),T1561(001,002)(E),T1489(E),T1499(001-004)(E),T1490(E),T1498(001,002)(E),T1496(E)                                                             \\
E: Enterprise, M: Mobile, I: Industrial Control Systems (ICS) &                                           &                                                                                                                                                                                       
\end{tblr}
}
\end{table*}

\par
Each tactic within the MITRE ATT\&CK is examined for every phase of the context-sharing platform. Consequently, we identify relevant techniques and sub-techniques associated with each phase of the context-sharing platform. Additionally, we establish a mapping between the MITRE techniques and sub-techniques and the primary components of the context-sharing platform. Identifying related MITRE techniques and sub-techniques for each component of the context-sharing platform helps identify potential mitigation methods outlined in the MITRE ATT\&CK. This, in turn, contributes to enhancing the security of the associated components. Given the substantial volume of MITRE techniques and sub-techniques identified for threat modelling in the context-sharing platforms, we present the mapped threats separately. This presentation is detailed in tables \ref{table:methodology:dataconsumer&data}, \ref{table:methodology:hardware&transmission}, and \ref{table:methodology:virtualisation}. In each table, rows represent various MITRE ATT\&CK tactics. Columns delineate the phases of the context-sharing platform, which are further categorized by the identified primary components of the context-sharing platform. 
\par
Creating a versatile threat modelling framework suitable for a wide range of IoT scenarios on a context-sharing platform requires a comprehensive analysis. Specifically, for each technique and sub-technique outlined in the MITRE ATT\&CK, we identify all the potential components of the context-sharing platform (e.g., hardware, transmission) to which a threat might be relevant. As a result, the mapping tables we have developed may contain common threats relevant to various context-sharing platform components. It is worth noting that the applicability of these threats depends on specific factors, such as the platform's architecture design, deployment model, and security assumptions. Consequently, only a subset of these threats will apply to each unique design or deployment.
\par
As an illustration of how threats are mapped in these tables, consider \textit{Endpoint Denial of Service (code T1499 (E))} threat. This threat has been recognized as a technique within the enterprise (E) domain under the \textit{Impact} tactic in the MITRE ATT\&CK. This technique comprises four sub-techniques: \textit{OS Exhaustion Flood (T1499.001), Service Exhaustion Flood (T1499.002), Application Exhaustion Flood (T1499.003)}, and \textit{Application or System Exploitation (T1499.004)}. For each sub-technique, we have conducted a thorough analysis of MITRE ATT\&CK's provided definitions and descriptions to pinpoint the primary targets of these sub-techniques. Consequently, we have delved into each sub-technique to determine which specific components within context-sharing platforms are relevant. For instance, consider the sub-technique \textit{OS Exhaustion Flood (T1499.001)}. This sub-technique can be associated with the OS/firmware component because its primary target is the OS. Additionally, it can be linked to the virtualisation component, given that virtualised OSs are a prime example of services offered by cloud service providers. Next, take \textit{Service exhaustion flood (T1499.002)}. This sub-technique aligns with the transmission component as attackers may aim at various network services provided by devices like routers. This sub-technique applies to the application component as well. For instance, think of an HTTP flood, where adversaries flood a web server with a high volume of HTTP requests to overwhelm it, which can also impact applications running atop the server. Additionally, it pertains to the virtualisation component since many network and application-level services operate within virtualised environments, such as the cloud. \textit{Application Exhaustion Flood (T1499.003)} is relevant to the transmission component since network devices can run certain applications. While some of these devices may not host public-facing applications, others do, especially in edge/fog computing scenarios where reducing access latency for critical applications is crucial. This sub-technique also applies to the application and virtualisation components for reasons similar to those previously discussed for other sub-techniques. Lastly, \textit{Application or System Exploitation (T1499.004)} encompasses known or zero-day vulnerabilities that can crash applications and systems. This threat is relevant to various components within context-sharing platforms, including transmission, hardware, application, and virtualisation. In Table~\ref{table:methodology:hardware&transmission} and Table~\ref{table:methodology:virtualisation}, we have documented these mappings within the \textit{Impact} tactic.
\par
When examining tables \ref{table:methodology:dataconsumer&data}, \ref{table:methodology:hardware&transmission}, and \ref{table:methodology:virtualisation}, our methodology offers an extensive compilation of threats (detailed based on technique and sub-techniques) based on the initial target of threats within scenarios applicable in context-sharing platforms. Also, our methodology can serve as a framework for systematically analysing the security of solutions proposed within the context-sharing platform domain.
\section{Evaluation}
\label{qualitative_evaluation}
%
%
\subsection{Academic Research}
For each academic research paper, we thoroughly examined the security components, their interactions with other system components, and the security-related features outlined in the paper. Subsequently, we used our methodology to assess potential threats that could be mitigated by leveraging these security features.
\par
Hassani et al.~\cite{hassani2018context} proposes a framework in which the security and communication manager module is responsible for analysing incoming messages to the systems and identifying anomalous patterns to mitigate threats such as Distributed Denial of Service (DDoS) attacks. In addition, it verifies the authorisation of users to access specific contextual information within the system. The context query engine manages the incoming queries to retrieve contextual data stored within the system. The context storage management system maintains a cache containing historical contextual information. This cache enhances response times when responding to queries by leveraging past context data. Finally, the context reasoning engine generates new contextual information by processing raw data. The framework proposed by Pereira et al.~\cite{pereira2022platform} uses the FIWARE context-sharing platform. This framework is structured in two layers: a) the platform middleware and b) the infrastructure middleware. The platform middleware consists of the main elements of the platform. The infrastructure middleware contains the underlying middleware services used by this framework, including the identity manager component, the authorisation component, the context broker, the IoT discovery component, the IoT gateway component, the context-aware engine component, the batch processor component, and the message broker. The security manager component is essential in safeguarding the data managed within this framework. The implemented security model relies on roles, access policies, OAuth protocol, and user information. It leverages functionalities the identity manager provides and authorisation components of the underlying middleware infrastructure. This component intercepts all incoming requests directed at the framework to ensure that only authenticated and authorised users can interact with the managed layers. It retrieves access tokens from users/applications embedded in the HTTP request header and forwards them to the identity manager component for validation. The security manager component also triggers the authorisation component within the underlying middleware to verify whether the user or application possesses the necessary permissions to execute the requested operation.
\par
Tables \ref{table:evaluation:contextextraction-papers} and \ref{table:evaluation:contextreasoning-papers} present the outcomes of applying our threat modelling with respect to the main phases of context-sharing solutions that fall into the `Academic Research' category. In our assessment, if a solution included security measures to mitigate even one of the identified threats based on our threat modelling framework within context-sharing platforms, we considered that work partially addressing the related threats. For instance, Hassani et al.~\cite{hassani2018context} had implemented a security mechanism against the DDoS threat, which falls under the impact tactics as depicted in Table \ref{table:evaluation:contextextraction-papers}. However, it is essential to note that there are more than 20 other threats within the impact tactics that target various components of the context-sharing platform, and these were not addressed in \cite{hassani2018context}. Nevertheless, since we applied a criterion of at least one security measure to determine partial coverage in the tables, the tables indicate partial coverage for the virtualisation, transmission, and application components within the impact tactics.
{
\scriptsize
\begin{table}[]
\centering
\caption{Evaluation of Academic Research - Context Extraction \& Distribution}
\label{table:evaluation:contextextraction-papers}
\resizebox{0.8\linewidth}{!}{%
\begin{tblr}{
  cells = {c},
  cell{1}{1} = {c=2}{},
  cell{1}{3} = {c=6}{},
  cell{1}{9} = {c=6}{},
  cell{2}{1} = {c=2}{},
  cell{3}{1} = {r=16}{},
  cell{19}{1} = {c=14}{},
  vlines,
  hline{1-3,19-20} = {-}{},
  hline{4-18} = {2-14}{},
}
Reference / Year                                                                                                                                    &                         & \cite{hassani2018context} / (2018) &    &    &    &    &    & \cite{pereira2022platform} / (2022) &    &    &    &    &    \\
Components                                                                                                                                          &                         & DC             & V  & H  & T  & O  & A  & DC             & V  & H  & T  & O  & A  \\
\begin{sideways}MITRE Tactics\end{sideways} &
Reconnaissance &
\scalebox{1.5}{\ensuremath \Circle} &
\scalebox{1.5}{\ensuremath \Circle} &
\scalebox{1.5}{\ensuremath \Circle} &
\scalebox{1.5}{\ensuremath \Circle} &
\scalebox{1.5}{\ensuremath \Circle} &
\scalebox{1.5}{\ensuremath \Circle} &
\scalebox{1.5}{\ensuremath \Circle} &
\scalebox{1.5}{\ensuremath \Circle} &
\scalebox{1.5}{\ensuremath \Circle} &
\scalebox{1.5}{\ensuremath \Circle} &
\scalebox{1.5}{\ensuremath \Circle} &
\scalebox{1.5}{\ensuremath \Circle} \\

& {Resource\\Development} &
\scalebox{1.5}{\ensuremath \Circle} &
\scalebox{1.5}{\ensuremath \Circle} &
\scalebox{1.5}{\ensuremath \Circle} &
\scalebox{1.5}{\ensuremath \Circle} &
\scalebox{1.5}{\ensuremath \Circle} &
\scalebox{1.5}{\ensuremath \Circle} &
\scalebox{1.5}{\ensuremath \Circle} &
\scalebox{1.5}{\ensuremath \Circle} &
\scalebox{1.5}{\ensuremath \Circle} &
\scalebox{1.5}{\ensuremath \Circle} &
\scalebox{1.5}{\ensuremath \Circle} &
\scalebox{1.5}{\ensuremath \Circle} \\

& {Initial\\Access} &
\scalebox{1.5}{\ensuremath \Circle} &
\scalebox{1.5}{\ensuremath \Circle} &
\scalebox{1.5}{\ensuremath \Circle} &
\scalebox{1.5}{\ensuremath \Circle} &
\scalebox{1.5}{\ensuremath \Circle} &
\scalebox{1.5}{\ensuremath \Circle} &
\scalebox{1.5}{\ensuremath \LEFTcircle} &
\scalebox{1.5}{\ensuremath \LEFTcircle} &
\scalebox{1.5}{\ensuremath \Circle} &
\scalebox{1.5}{\ensuremath \LEFTcircle} &
\scalebox{1.5}{\ensuremath \LEFTcircle} &
\scalebox{1.5}{\ensuremath \LEFTcircle} \\

& Execution &
\scalebox{1.5}{\ensuremath \Circle} &
\scalebox{1.5}{\ensuremath \Circle} &
\scalebox{1.5}{\ensuremath \Circle} &
\scalebox{1.5}{\ensuremath \Circle} &
\scalebox{1.5}{\ensuremath \Circle} &
\scalebox{1.5}{\ensuremath \Circle} &
\scalebox{1.5}{\ensuremath \Circle} &
\scalebox{1.5}{\ensuremath \Circle} &
\scalebox{1.5}{\ensuremath \LEFTcircle} &
\scalebox{1.5}{\ensuremath \LEFTcircle} &
\scalebox{1.5}{\ensuremath \LEFTcircle} &
\scalebox{1.5}{\ensuremath \LEFTcircle} \\

& Persistence &
\scalebox{1.5}{\ensuremath \Circle} &
\scalebox{1.5}{\ensuremath \Circle} &
\scalebox{1.5}{\ensuremath \Circle} &
\scalebox{1.5}{\ensuremath \Circle} &
\scalebox{1.5}{\ensuremath \Circle} &
\scalebox{1.5}{\ensuremath \Circle} &
\scalebox{1.5}{\ensuremath \LEFTcircle} &
\scalebox{1.5}{\ensuremath \LEFTcircle} &
\scalebox{1.5}{\ensuremath \Circle} &
\scalebox{1.5}{\ensuremath \LEFTcircle} &
\scalebox{1.5}{\ensuremath \LEFTcircle} &
\scalebox{1.5}{\ensuremath \LEFTcircle} \\

& {Privilege\\Escalation} &
\scalebox{1.5}{\ensuremath \Circle} &
\scalebox{1.5}{\ensuremath \Circle} &
\scalebox{1.5}{\ensuremath \Circle} &
\scalebox{1.5}{\ensuremath \Circle} &
\scalebox{1.5}{\ensuremath \Circle} &
\scalebox{1.5}{\ensuremath \Circle} &
\scalebox{1.5}{\ensuremath \LEFTcircle} &
\scalebox{1.5}{\ensuremath \LEFTcircle} &
\scalebox{1.5}{\ensuremath \Circle} &
\scalebox{1.5}{\ensuremath \Circle} &
\scalebox{1.5}{\ensuremath \LEFTcircle} &
\scalebox{1.5}{\ensuremath \LEFTcircle} \\

& {Defense\\Evasion} &
\scalebox{1.5}{\ensuremath \Circle} &
\scalebox{1.5}{\ensuremath \Circle} &
\scalebox{1.5}{\ensuremath \Circle} &
\scalebox{1.5}{\ensuremath \Circle} &
\scalebox{1.5}{\ensuremath \Circle} &
\scalebox{1.5}{\ensuremath \Circle} &
\scalebox{1.5}{\ensuremath \LEFTcircle} &
\scalebox{1.5}{\ensuremath \LEFTcircle} &
\scalebox{1.5}{\ensuremath \LEFTcircle} &
\scalebox{1.5}{\ensuremath \LEFTcircle} &
\scalebox{1.5}{\ensuremath \LEFTcircle} &
\scalebox{1.5}{\ensuremath \LEFTcircle} \\

& {Credential\\Access} &
\scalebox{1.5}{\ensuremath \Circle} &
\scalebox{1.5}{\ensuremath \Circle} &
\scalebox{1.5}{\ensuremath \Circle} &
\scalebox{1.5}{\ensuremath \Circle} &
\scalebox{1.5}{\ensuremath \Circle} &
\scalebox{1.5}{\ensuremath \Circle} &
\scalebox{1.5}{\ensuremath \LEFTcircle} &
\scalebox{1.5}{\ensuremath \LEFTcircle} &
\scalebox{1.5}{\ensuremath \LEFTcircle} &
\scalebox{1.5}{\ensuremath \LEFTcircle} &
\scalebox{1.5}{\ensuremath \LEFTcircle} &
\scalebox{1.5}{\ensuremath \LEFTcircle} \\

& Discovery &
\scalebox{1.5}{\ensuremath \Circle} &
\scalebox{1.5}{\ensuremath \Circle} &
\scalebox{1.5}{\ensuremath \Circle} &
\scalebox{1.5}{\ensuremath \Circle} &
\scalebox{1.5}{\ensuremath \Circle} &
\scalebox{1.5}{\ensuremath \Circle} &
\scalebox{1.5}{\ensuremath \Circle} &
\scalebox{1.5}{\ensuremath \Circle} &
\scalebox{1.5}{\ensuremath \Circle} &
\scalebox{1.5}{\ensuremath \Circle} &
\scalebox{1.5}{\ensuremath \Circle} &
\scalebox{1.5}{\ensuremath \Circle} \\

& {Lateral\\Movement} &
\scalebox{1.5}{\ensuremath \Circle} &
\scalebox{1.5}{\ensuremath \Circle} &
\scalebox{1.5}{\ensuremath \Circle} &
\scalebox{1.5}{\ensuremath \Circle} &
\scalebox{1.5}{\ensuremath \Circle} &
\scalebox{1.5}{\ensuremath \Circle} &
\scalebox{1.5}{\ensuremath \Circle} &
\scalebox{1.5}{\ensuremath \LEFTcircle} &
\scalebox{1.5}{\ensuremath \LEFTcircle} &
\scalebox{1.5}{\ensuremath \LEFTcircle} &
\scalebox{1.5}{\ensuremath \LEFTcircle} &
\scalebox{1.5}{\ensuremath \LEFTcircle} \\

& Collection &
\scalebox{1.5}{\ensuremath \LEFTcircle} &
\scalebox{1.5}{\ensuremath \Circle} &
\scalebox{1.5}{\ensuremath \Circle} &
\scalebox{1.5}{\ensuremath \Circle} &
\scalebox{1.5}{\ensuremath \Circle} &
\scalebox{1.5}{\ensuremath \Circle} &
\scalebox{1.5}{\ensuremath \LEFTcircle} &
\scalebox{1.5}{\ensuremath \LEFTcircle} &
\scalebox{1.5}{\ensuremath \LEFTcircle} &
\scalebox{1.5}{\ensuremath \LEFTcircle} &
\scalebox{1.5}{\ensuremath \LEFTcircle} &
\scalebox{1.5}{\ensuremath \LEFTcircle} \\

& {Command\\Control} &
\scalebox{1.5}{\ensuremath \Circle} &
\scalebox{1.5}{\ensuremath \Circle} &
\scalebox{1.5}{\ensuremath \Circle} &
\scalebox{1.5}{\ensuremath \Circle} &
\scalebox{1.5}{\ensuremath \Circle} &
\scalebox{1.5}{\ensuremath \Circle} &
\scalebox{1.5}{\ensuremath \Circle} &
\scalebox{1.5}{\ensuremath \Circle} &
\scalebox{1.5}{\ensuremath \Circle} &
\scalebox{1.5}{\ensuremath \Circle} &
\scalebox{1.5}{\ensuremath \Circle} &
\scalebox{1.5}{\ensuremath \Circle} \\

& Exfiltration &
\scalebox{1.5}{\ensuremath \Circle} &
\scalebox{1.5}{\ensuremath \Circle} &
\scalebox{1.5}{\ensuremath \Circle} &
\scalebox{1.5}{\ensuremath \Circle} &
\scalebox{1.5}{\ensuremath \Circle} &
\scalebox{1.5}{\ensuremath \Circle} &
\scalebox{1.5}{\ensuremath \Circle} &
\scalebox{1.5}{\ensuremath \Circle} &
\scalebox{1.5}{\ensuremath \Circle} &
\scalebox{1.5}{\ensuremath \Circle} &
\scalebox{1.5}{\ensuremath \Circle} &
\scalebox{1.5}{\ensuremath \Circle} \\

& {Inhibit\\Response} &
\scalebox{1.5}{\ensuremath \Circle} &
\scalebox{1.5}{\ensuremath \Circle} &
\scalebox{1.5}{\ensuremath \Circle} &
\scalebox{1.5}{\ensuremath \Circle} &
\scalebox{1.5}{\ensuremath \Circle} &
\scalebox{1.5}{\ensuremath \Circle} &
\scalebox{1.5}{\ensuremath \Circle} &
\scalebox{1.5}{\ensuremath \Circle} &
\scalebox{1.5}{\ensuremath \Circle} &
\scalebox{1.5}{\ensuremath \Circle} &
\scalebox{1.5}{\ensuremath \Circle} &
\scalebox{1.5}{\ensuremath \Circle} \\

& {Impair\\Control} &
\scalebox{1.5}{\ensuremath \Circle} &
\scalebox{1.5}{\ensuremath \Circle} &
\scalebox{1.5}{\ensuremath \Circle} &
\scalebox{1.5}{\ensuremath \Circle} &
\scalebox{1.5}{\ensuremath \Circle} &
\scalebox{1.5}{\ensuremath \Circle} &
\scalebox{1.5}{\ensuremath \Circle} &
\scalebox{1.5}{\ensuremath \Circle} &
\scalebox{1.5}{\ensuremath \Circle} &
\scalebox{1.5}{\ensuremath \Circle} &
\scalebox{1.5}{\ensuremath \Circle} &
\scalebox{1.5}{\ensuremath \Circle} \\

& Impact &
\scalebox{1.5}{\ensuremath \Circle} &
\scalebox{1.5}{\ensuremath \LEFTcircle} &
\scalebox{1.5}{\ensuremath \Circle} &
\scalebox{1.5}{\ensuremath \LEFTcircle} &
\scalebox{1.5}{\ensuremath \Circle} &
\scalebox{1.5}{\ensuremath \LEFTcircle} &
\scalebox{1.5}{\ensuremath \Circle} &
\scalebox{1.5}{\ensuremath \Circle} &
\scalebox{1.5}{\ensuremath \Circle} &
\scalebox{1.5}{\ensuremath \Circle} &
\scalebox{1.5}{\ensuremath \Circle} &
\scalebox{1.5}{\ensuremath \Circle} \\

{\scalebox{1.5}{\ensuremath \CIRCLE}: Full Cover, \scalebox{1.5}{\ensuremath \LEFTcircle}: Partial Cover, \scalebox{1.5}{\ensuremath \Circle}: Does Not Cover\\DC: Data Consumer, V: Virtualisation, H: Hardware, T: Transmission,\\O: OS/Firmware, A: Application} & & & & & & & & & & & & \\
   
\end{tblr}
}
\end{table}
\vspace{-0.25cm}
}

{
\scriptsize
\begin{table} []
\centering
\caption{Evaluation of Academic Research - Context Modelling \& Reasoning}
\label{table:evaluation:contextreasoning-papers}
\resizebox{0.6\columnwidth}{!}{
\renewcommand{\arraystretch}{0.8} 
\begin{tblr}{
  cells = {c},
  cell{1}{1} = {c=2}{},
  cell{2}{1} = {r=16}{},
  cell{18}{1} = {c=4}{},
  vlines,
  hline{1-2,18-19} = {-}{},
  hline{3-17} = {2-4}{}, 
}
References / Year                         &    & \cite{hassani2018context} / (2018) & \cite{pereira2022platform} / (2022) \\
\begin{sideways}\scriptsize MITRE Tactics\end{sideways} & Reconnaissance          & \scalebox{1.2}{\ensuremath \Circle}              & \scalebox{1.2}{\ensuremath \Circle}              \\
                                            & {Resource\\Development} &  \scalebox{1.2}{\ensuremath \Circle}              & \scalebox{1.2}{\ensuremath \Circle}                \\
                                            & {Initial\\Access}       &  \scalebox{1.2}{\ensuremath \Circle}              & \scalebox{1.2}{\ensuremath \Circle} \\
                                            & Execution               &  \scalebox{1.2}{\ensuremath \Circle}              & \scalebox{1.2}{\ensuremath \Circle}                   \\
                                            & Persistence             &  \scalebox{1.2}{\ensuremath \Circle}              & \scalebox{1.2}{\ensuremath \Circle}                \\
                                            & {Privilage\\Escalation} &  \scalebox{1.2}{\ensuremath \Circle}              & \scalebox{1.2}{\ensuremath \Circle}                \\
                                            & {Defense\\Evasion}      &  \scalebox{1.2}{\ensuremath \Circle}              & \scalebox{1.2}{\ensuremath \LEFTcircle}                            \\
                                            & {Credential\\Access}   &  \scalebox{1.2}{\ensuremath \Circle}              & \scalebox{1.2}{\ensuremath \Circle}           \\
                                            & Discovery               &  \scalebox{1.2}{\ensuremath \Circle}             & \scalebox{1.2}{\ensuremath \Circle}                          \\
                                            & {Lateral\\Movement}     &  \scalebox{1.2}{\ensuremath \Circle}              & \scalebox{1.2}{\ensuremath \Circle}                \\
                                            & Collection             &  \scalebox{1.2}{\ensuremath \LEFTcircle}              & \scalebox{1.2}{\ensuremath \LEFTcircle}                            \\
                                            & {Command\\Control}     & \scalebox{1.2}{\ensuremath \Circle}              & \scalebox{1.2}{\ensuremath \Circle}                \\
                                            & Exfilteration           &  \scalebox{1.2}{\ensuremath \Circle}              & \scalebox{1.2}{\ensuremath \Circle}                     \\
                                            & {Inhibit \\ Response}                        & \scalebox{1.2}{\ensuremath \Circle}              & \scalebox{1.2}{\ensuremath \Circle}                        \\
                                            & {Impair\\Control}       &  \scalebox{1.2}{\ensuremath \Circle}              & \scalebox{1.2}{\ensuremath \Circle}                     \\
                                            & Impact                  &  \scalebox{1.2}{\ensuremath \Circle}             & \scalebox{1.2}{\ensuremath \Circle}                     \\
\scalebox{1.2}{\ensuremath \CIRCLE}: Full Cover, \scalebox{1.2}{\ensuremath \LEFTcircle}: Partial Cover,  \scalebox{1.2}{\ensuremath \Circle}: Does Not Cover  &                         &                        &                 
\end{tblr}
}
\end{table}
}
\subsection{Industry Projects}
In this section, we study two industry projects and apply our methodology to assess their security considerations. To analyse industry projects, we focus on projects' publicly available deliverables\footnote{The link to each project and its associated deliverables are provided in \S\ref{background}}. For each project, we have examined the security components, their interactions with other system components, and the security-related features outlined in the project. Subsequently, we used our methodology to assess the potential threats that could be mitigated by leveraging these security features. Given the comprehensive content of security-related deliverables in projects, which often extends across multiple deliverables, we only provide a summary of the key security features in this section.
\par
Within the IDIRA project, information exchange primarily relies on Simple Object Access Protocol (SOAP) Web Services for synchronous communication between software components, both internally and with external systems, as well as REST web services. IDIRA ensures the security of web services and implements a process to authenticate and encrypt communications among all connected software systems.
\par
In the SEMIoTICs project, software components are organized into three layers: the field layer, SDN/NFV Orchestration layer, and application orchestration layer. Each layer incorporates a security manager to address security requirements specific to that layer. These components collaborate to authenticate users and manage their identities, oversee the identities of other entities such as sensors, enforce access controls for privacy-sensitive information within applications, and ensure end-to-end secure networking capabilities. SEMIoTICS components leverage authentication services the security manager provides on the back end. Applications needing authentication services must register as OAuth2 clients with agile security to redirect users to the authentication endpoint. Additionally, the Web of Things protocol facilitates interoperability by offering an abstract semantic interface over widely used transport layers like HTTP Rest APIs, CoAP, and MQTT, which support authentication and data encryption. Also, SEMIoTiCs use attribute-based identity management. For encryption, keys are regularly regenerated to enhance security. This measure ensures that any compromised key is only usable until the next key regeneration cycle. In the SDN/NFV layer, the security manager provides authentication and accounting services to entities interacting with the controller. Its primary function is to support authentication and accounting services for tenant administration and application assignment, issuing tokens for rapid authentication during runtime.
\par
Tables \ref{table:evaluation:contextextraction-industry} and \ref{table:evaluation:contextereasoning-industry} present the outcomes of applying our threat modelling, focusing on the context-sharing and distribution as well as context modelling and reasoning phases in industry projects, respectively.

{
\scriptsize
\begin{table}
\centering
\caption{Evaluation of Industry Projects - Context Extraction \& Context Distribution}
\label{table:evaluation:contextextraction-industry}
\resizebox{0.7\linewidth}{!}{%
\begin{tblr}{
  cells = {c},
  cell{1}{1} = {c=2}{},
  cell{1}{3} = {c=6}{},
  cell{1}{9} = {c=6}{},
  cell{2}{1} = {c=2}{},
  cell{3}{1} = {r=16}{},
  cell{19}{1} = {c=14}{},
  vlines,
  hline{1-3,19-20} = {-}{},
  hline{4-18} = {2-14}{},
}
Reference / Year                                                                                                                                                 &                         & IDIRA / (2011-2015) &       &       &       &       &       & SEMIOTICS / (2018-2020) &       &       &       &       &       \\
Components                                                                                                                                                       &                         & DC                   & V     & H     & T     & O     & A     & DC                       & V     & H     & T     & O     & A     \\
\begin{sideways}MITRE Tactics\end{sideways}                                                                                                                      & Reconnaissance          & \scalebox{1.5}{\ensuremath \Circle}               & \scalebox{1.5}{\ensuremath \Circle} & \scalebox{1.5}{\ensuremath \Circle} & \scalebox{1.5}{\ensuremath \Circle} & \scalebox{1.5}{\ensuremath \Circle} & \scalebox{1.5}{\ensuremath \Circle} & \scalebox{1.5}{\ensuremath \Circle}                   & \scalebox{1.5}{\ensuremath \Circle} & \scalebox{1.5}{\ensuremath \Circle} & \scalebox{1.5}{\ensuremath \Circle} & \scalebox{1.5}{\ensuremath \Circle} & \scalebox{1.5}{\ensuremath \Circle} \\  & {Resource\\Development} & \scalebox{1.5}{\ensuremath \Circle}               & \scalebox{1.5}{\ensuremath \Circle} & \scalebox{1.5}{\ensuremath \Circle} & \scalebox{1.5}{\ensuremath \Circle} & \scalebox{1.5}{\ensuremath \Circle} & \scalebox{1.5}{\ensuremath \Circle} & \scalebox{1.5}{\ensuremath \Circle}                   & \scalebox{1.5}{\ensuremath \Circle} & \scalebox{1.5}{\ensuremath \Circle} & \scalebox{1.5}{\ensuremath \Circle} & \scalebox{1.5}{\ensuremath \Circle} & \scalebox{1.5}{\ensuremath \Circle} \\  & {Initial\\Access}       & \scalebox{1.5}{\ensuremath \Circle}               & \scalebox{1.5}{\ensuremath \Circle} & \scalebox{1.5}{\ensuremath \Circle} & \scalebox{1.5}{\ensuremath \Circle} & \scalebox{1.5}{\ensuremath \Circle} & \scalebox{1.5}{\ensuremath \LEFTcircle}  & \scalebox{1.5}{\ensuremath \Circle}                   & \scalebox{1.5}{\ensuremath \LEFTcircle}  & \scalebox{1.5}{\ensuremath \Circle} & \scalebox{1.5}{\ensuremath \LEFTcircle}  & \scalebox{1.5}{\ensuremath \Circle} & \scalebox{1.5}{\ensuremath \LEFTcircle}  \\  & Execution               & \scalebox{1.5}{\ensuremath \Circle}               & \scalebox{1.5}{\ensuremath \Circle} & \scalebox{1.5}{\ensuremath \Circle} & \scalebox{1.5}{\ensuremath \Circle} & \scalebox{1.5}{\ensuremath \Circle} & \scalebox{1.5}{\ensuremath \Circle} & \scalebox{1.5}{\ensuremath \Circle}                   & \scalebox{1.5}{\ensuremath \LEFTcircle}  & \scalebox{1.5}{\ensuremath \LEFTcircle}  & \scalebox{1.5}{\ensuremath \LEFTcircle}  & \scalebox{1.5}{\ensuremath \Circle} & \scalebox{1.5}{\ensuremath \LEFTcircle}  \\   & Persistence             & \scalebox{1.5}{\ensuremath \Circle}               & \scalebox{1.5}{\ensuremath \Circle} & \scalebox{1.5}{\ensuremath \Circle} & \scalebox{1.5}{\ensuremath \Circle} & \scalebox{1.5}{\ensuremath \Circle} & \scalebox{1.5}{\ensuremath \Circle} & \scalebox{1.5}{\ensuremath \Circle}                   & \scalebox{1.5}{\ensuremath \Circle} & \scalebox{1.5}{\ensuremath \LEFTcircle}  & \scalebox{1.5}{\ensuremath \Circle} & \scalebox{1.5}{\ensuremath \Circle} & \scalebox{1.5}{\ensuremath \LEFTcircle}  \\   & {Privilage\\Escalation} & \scalebox{1.5}{\ensuremath \Circle}               & \scalebox{1.5}{\ensuremath \Circle} & \scalebox{1.5}{\ensuremath \Circle} & \scalebox{1.5}{\ensuremath \Circle} & \scalebox{1.5}{\ensuremath \Circle} & \scalebox{1.5}{\ensuremath \Circle} & \scalebox{1.5}{\ensuremath \LEFTcircle}                    & \scalebox{1.5}{\ensuremath \Circle} & \scalebox{1.5}{\ensuremath \Circle} & \scalebox{1.5}{\ensuremath \Circle} & \scalebox{1.5}{\ensuremath \Circle} & \scalebox{1.5}{\ensuremath \LEFTcircle}  \\   & {Defense\\Evasion}      & \scalebox{1.5}{\ensuremath \Circle}               & \scalebox{1.5}{\ensuremath \Circle} & \scalebox{1.5}{\ensuremath \Circle} & \scalebox{1.5}{\ensuremath \Circle} & \scalebox{1.5}{\ensuremath \Circle} & \scalebox{1.5}{\ensuremath \Circle} & \scalebox{1.5}{\ensuremath \LEFTcircle}                    & \scalebox{1.5}{\ensuremath \Circle} & \scalebox{1.5}{\ensuremath \Circle} & \scalebox{1.5}{\ensuremath \Circle} & \scalebox{1.5}{\ensuremath \Circle} & \scalebox{1.5}{\ensuremath \LEFTcircle}  \\  & {Credential\\Access}    & \scalebox{1.5}{\ensuremath \LEFTcircle}                & \scalebox{1.5}{\ensuremath \LEFTcircle}  & \scalebox{1.5}{\ensuremath \Circle} & \scalebox{1.5}{\ensuremath \LEFTcircle}  & \scalebox{1.5}{\ensuremath \Circle} & \scalebox{1.5}{\ensuremath \LEFTcircle}  & \scalebox{1.5}{\ensuremath \LEFTcircle}                    & \scalebox{1.5}{\ensuremath \LEFTcircle}  & \scalebox{1.5}{\ensuremath \LEFTcircle}  & \scalebox{1.5}{\ensuremath \LEFTcircle}  & \scalebox{1.5}{\ensuremath \Circle} & \scalebox{1.5}{\ensuremath \LEFTcircle}  \\  & Discovery               & \scalebox{1.5}{\ensuremath \Circle}               & \scalebox{1.5}{\ensuremath \Circle} & \scalebox{1.5}{\ensuremath \Circle} & \scalebox{1.5}{\ensuremath \LEFTcircle}  & \scalebox{1.5}{\ensuremath \Circle} & \scalebox{1.5}{\ensuremath \LEFTcircle}  & \scalebox{1.5}{\ensuremath \Circle}                   & \scalebox{1.5}{\ensuremath \Circle} & \scalebox{1.5}{\ensuremath \Circle} & \scalebox{1.5}{\ensuremath \LEFTcircle}  & \scalebox{1.5}{\ensuremath \Circle} & \scalebox{1.5}{\ensuremath \LEFTcircle}  \\   & {Lateral\\Movement}     & \scalebox{1.5}{\ensuremath \Circle}               & \scalebox{1.5}{\ensuremath \Circle} & \scalebox{1.5}{\ensuremath \Circle} & \scalebox{1.5}{\ensuremath \Circle} & \scalebox{1.5}{\ensuremath \Circle} & \scalebox{1.5}{\ensuremath \LEFTcircle}  & \scalebox{1.5}{\ensuremath \Circle}                   & \scalebox{1.5}{\ensuremath \LEFTcircle}  & \scalebox{1.5}{\ensuremath \Circle} & \scalebox{1.5}{\ensuremath \LEFTcircle}  & \scalebox{1.5}{\ensuremath \Circle} & \scalebox{1.5}{\ensuremath \LEFTcircle}  \\  & Collection              & \scalebox{1.5}{\ensuremath \LEFTcircle}                & \scalebox{1.5}{\ensuremath \Circle} & \scalebox{1.5}{\ensuremath \Circle} & \scalebox{1.5}{\ensuremath \LEFTcircle}  & \scalebox{1.5}{\ensuremath \Circle} & \scalebox{1.5}{\ensuremath \Circle} & \scalebox{1.5}{\ensuremath \LEFTcircle}                    & \scalebox{1.5}{\ensuremath \LEFTcircle}  & \scalebox{1.5}{\ensuremath \Circle} & \scalebox{1.5}{\ensuremath \LEFTcircle}  & \scalebox{1.5}{\ensuremath \Circle} & \scalebox{1.5}{\ensuremath \Circle} \\ & {Command \\Control}     & \scalebox{1.5}{\ensuremath \Circle}               & \scalebox{1.5}{\ensuremath \Circle} & \scalebox{1.5}{\ensuremath \Circle} & \scalebox{1.5}{\ensuremath \Circle} & \scalebox{1.5}{\ensuremath \Circle} & \scalebox{1.5}{\ensuremath \Circle} & \scalebox{1.5}{\ensuremath \Circle}                   & \scalebox{1.5}{\ensuremath \LEFTcircle}  & \scalebox{1.5}{\ensuremath \Circle} & \scalebox{1.5}{\ensuremath \LEFTcircle}  & \scalebox{1.5}{\ensuremath \Circle} & \scalebox{1.5}{\ensuremath \LEFTcircle}  \\   & Exfilteration           & \scalebox{1.5}{\ensuremath \Circle}               & \scalebox{1.5}{\ensuremath \Circle} & \scalebox{1.5}{\ensuremath \Circle} & \scalebox{1.5}{\ensuremath \Circle} & \scalebox{1.5}{\ensuremath \Circle} & \scalebox{1.5}{\ensuremath \Circle} & \scalebox{1.5}{\ensuremath \Circle}                   & \scalebox{1.5}{\ensuremath \Circle} & \scalebox{1.5}{\ensuremath \Circle} & \scalebox{1.5}{\ensuremath \Circle} & \scalebox{1.5}{\ensuremath \Circle} & \scalebox{1.5}{\ensuremath \Circle} \\  & {Inhibit\\Response}     & \scalebox{1.5}{\ensuremath \Circle}               & \scalebox{1.5}{\ensuremath \Circle} & \scalebox{1.5}{\ensuremath \Circle} & \scalebox{1.5}{\ensuremath \Circle} & \scalebox{1.5}{\ensuremath \Circle} & \scalebox{1.5}{\ensuremath \Circle} & \scalebox{1.5}{\ensuremath \Circle}                   & \scalebox{1.5}{\ensuremath \Circle} & \scalebox{1.5}{\ensuremath \Circle} & \scalebox{1.5}{\ensuremath \Circle} & \scalebox{1.5}{\ensuremath \Circle} & \scalebox{1.5}{\ensuremath \Circle} \\ & {Impair\\Control}       & \scalebox{1.5}{\ensuremath \Circle}               & \scalebox{1.5}{\ensuremath \Circle} & \scalebox{1.5}{\ensuremath \Circle} & \scalebox{1.5}{\ensuremath \Circle} & \scalebox{1.5}{\ensuremath \Circle} & \scalebox{1.5}{\ensuremath \Circle} & \scalebox{1.5}{\ensuremath \Circle}                   & \scalebox{1.5}{\ensuremath \Circle} & \scalebox{1.5}{\ensuremath \Circle} & \scalebox{1.5}{\ensuremath \Circle} & \scalebox{1.5}{\ensuremath \Circle} & \scalebox{1.5}{\ensuremath \Circle} \\  & Impact                  & \scalebox{1.5}{\ensuremath \LEFTcircle}                & \scalebox{1.5}{\ensuremath \Circle} & \scalebox{1.5}{\ensuremath \Circle} & \scalebox{1.5}{\ensuremath \Circle} & \scalebox{1.5}{\ensuremath \Circle} & \scalebox{1.5}{\ensuremath \Circle} & \scalebox{1.5}{\ensuremath \LEFTcircle}                    & \scalebox{1.5}{\ensuremath \LEFTcircle}  & \scalebox{1.5}{\ensuremath \Circle} & \scalebox{1.5}{\ensuremath \LEFTcircle}  & \scalebox{1.5}{\ensuremath \Circle} & \scalebox{1.5}{\ensuremath \LEFTcircle}  \\
{\scalebox{1.5}{\ensuremath \CIRCLE}: Full Cover, \scalebox{1.5}{\ensuremath \LEFTcircle}: Partial Cover,  \scalebox{1.5}{\ensuremath \Circle}: Does Not Cover\\DC: Data Consumer, V: Virtualisation, H: Hardware, T:Transmission,\\ O: OS/Firmware, A: Application}  &                         &                     &       &       &       &       &       &                         &       &       &       &       &       
\end{tblr}
}
\end{table}
}


{
\scriptsize
\begin{table}[t]
\centering
\caption{Evaluation of Industry Projects - Context Modeling \& Context Reasoning}
\label{table:evaluation:contextereasoning-industry}
\resizebox{0.7\linewidth}{!}{%
\begin{tblr}{
  row{odd} = {c},
  row{4} = {c},
  row{6} = {c},
  row{8} = {c},
  row{10} = {c},
  row{12} = {c},
  row{14} = {c},
  row{16} = {c},
  cell{1}{1} = {c=2}{},
  cell{2}{1} = {r=16}{},
  cell{2}{2} = {c},
  cell{2}{3} = {c},
  cell{2}{4} = {c},
  cell{18}{1} = {c=4}{},
  vlines,
  hline{1-2,18-19} = {-}{},
  hline{3-17} = {2-4}{},
}
References / Year                           &                         & IDIRA / (2011-2015) & SEMIOTICS / (2018-2020) \\
\begin{sideways}MITRE Tactics\end{sideways} & Reconnaissance          & \scalebox{1.5}{\ensuremath \Circle}               & \scalebox{1.5}{\ensuremath \Circle}                   \\
                                            & {Resource\\Development} & \scalebox{1.5}{\ensuremath \Circle}               & \scalebox{1.5}{\ensuremath \Circle}                   \\
                                            & {Initial\\Access}       & \scalebox{1.5}{\ensuremath \Circle}               & \scalebox{1.5}{\ensuremath \Circle}                   \\
                                            & Execution               & \scalebox{1.5}{\ensuremath \Circle}               & \scalebox{1.5}{\ensuremath \Circle}                   \\
                                            & Persistence             & \scalebox{1.5}{\ensuremath \Circle}               & \scalebox{1.5}{\ensuremath \Circle}                   \\
                                            & {Privilage\\Escalation} & \scalebox{1.5}{\ensuremath \Circle}               & \scalebox{1.5}{\ensuremath \Circle}                   \\
                                            & {Defense\\Evasion}      & \scalebox{1.5}{\ensuremath \Circle}               & \scalebox{1.5}{\ensuremath \Circle}                   \\
                                            & {Credential\\Access}    & \scalebox{1.5}{\ensuremath \Circle}               & \scalebox{1.5}{\ensuremath \LEFTcircle}                    \\
                                            & Discovery               & \scalebox{1.5}{\ensuremath \Circle}               & \scalebox{1.5}{\ensuremath \Circle}                   \\
                                            & {Lateral\\Movement}     & \scalebox{1.5}{\ensuremath \Circle}               & \scalebox{1.5}{\ensuremath \Circle}                   \\
                                            & Collection              & \scalebox{1.5}{\ensuremath \LEFTcircle}                & \scalebox{1.5}{\ensuremath \LEFTcircle}                    \\
                                            & {Command \\Control}     & \scalebox{1.5}{\ensuremath \Circle}               & \scalebox{1.5}{\ensuremath \Circle}                   \\
                                            & Exfilteration           & \scalebox{1.5}{\ensuremath \Circle}               & \scalebox{1.5}{\ensuremath \Circle}                   \\
                                            & {Inhibit\\Response}     & \scalebox{1.5}{\ensuremath \Circle}               & \scalebox{1.5}{\ensuremath \Circle}                   \\
                                            & {Impair\\Control}       & \scalebox{1.5}{\ensuremath \Circle}               & \scalebox{1.5}{\ensuremath \Circle}                   \\
                                            & Impact                  & \scalebox{1.5}{\ensuremath \LEFTcircle}                & \scalebox{1.5}{\ensuremath \LEFTcircle}                    \\
                                         \scalebox{1.5}{\ensuremath \CIRCLE}: Full Cover, \scalebox{1.5}{\ensuremath \LEFTcircle}: Partial Cover,  \scalebox{1.5}{\ensuremath \Circle}: Does Not Cover  &                         &                        &        
\end{tblr}
}
\end{table}
}
\subsection{Discussion \& Open-Source Tool}
While most academic research in context-sharing platforms acknowledges security and privacy as a critical requirement, only a few studies incorporate explicit security considerations. This underscores the critical importance of security in context-sharing solutions, which serve as foundational layers for developing and deploying diverse IoT applications. 
\par
Our findings indicate that context-sharing solutions' context extraction and distribution phases have garnered more attention in academic research and industry projects. Conversely, the context modelling and reasoning phases have been less emphasized. Data, whether raw or stored, is central to the context modelling and reasoning phases, necessitating greater emphasis on its security considerations. Regarding components, data consumers, virtualisation, transmission, and applications have received more security focus, while data, operating systems, and hardware have comparatively received less attention.
\par
Our examination of context-sharing solutions against ATT\&CK reveals that academic research places greater emphasis on collection, credential access, and defence evasion tactics, while reconnaissance, resource development, discovery, command \& control, exfiltration, inhibit response, and impair control receive less attention. Industry projects mirror this trend in prioritising tactics, yet they assign more importance to the collection, credential access, and impact tactics. Therefore, as a crucial research gap, it is imperative to redesign the security mechanisms of context-sharing solutions to consider the least prioritized threats or implement relevant mitigation techniques to address them.
\par
We observed that neither academic research nor industry projects effectively mitigate all identified threats within any category and lack a holistic and systematic approach to mitigate security risks applicable to IoT context-sharing platforms. While overall security considerations in academic research are limited, industry projects tend to have more elaborate security considerations. For instance, Table \ref{table:methodology:dataconsumer&data} highlights fifteen threats within the collection tactic identified for the data consumer component. Both Pereira et al. \cite{pereira2022platform} and SEMIoTICs only partially address the collection tactic of the data consumer, as indicated in Table \ref{table:evaluation:contextextraction-papers} and Table \ref{table:evaluation:contextextraction-industry}. When applying our threat modelling methodology to academic research and industry projects, we consistently used policy at least once, as discussed earlier. In this instance, SEMIoTICs addressed more of the overall fifteen threats. This trend was observed across other tactics and components, indicating that industry projects tend to have more comprehensive overall security considerations.
\par
Our threat modelling framework is designed for generic CSP rather than tailored for IoT application scenarios. We used this and developed an open-source threat analysis tool for context-sharing platforms available from \url{https://mgoudarzi90.github.io/ThreatModelling_IoT_CSP}. This tool will enable the evaluation of the security of different CSPs by providing the relevant threats (covering all relevant techniques and sub-techniques) and generating a report. Specifically, the detailed list of techniques and sub-techniques developed against a generic IoT CSP can facilitate a complete threat modelling of any specific system. Furthermore, since we mapped ATT\&CK tactics and threats to various components of CSP, relevant potential mitigation techniques already available in MITRE can be considered as potential mitigation techniques for each component of CSP. We plan to integrate the mitigation recommendations into future tool versions and improve the report generated. We note that selecting specific mitigation techniques for each component and tactic depends on the specific CSP or IoT application.
%
%
%
%
%
\section{Conclusion}
\label{conclusion_futureWork}

We take the initial steps towards developing a comprehensive threat modelling of context-sharing platforms to systematically analyse potential threats within such systems. We have identified the main components of generic IoT context-sharing platforms by analysing several key industry projects and relevant research papers. We have employed the MITRE ATT\&CK to methodically analyse the complete range of threats that adversaries might employ to target a generic context-sharing solution listing all applicable techniques and sub-techniques. Our analysis yields a detailed threat model that can facilitate the security assessment of existing IoT context-sharing platforms and enhance future solutions' security through a 'secure-by-design' approach. We have made an open-source tool available to aid future projects and research in achieving this goal. In future, we will expand our analysis and evaluate the design and deployment of ongoing industry projects such as FIWARE. Complementing our open-source tool, we also plan to develop guidelines for developing secure context-sharing platforms for IoT. 

\bibliographystyle{ieeetr}
\bibliography{ref}

\begin{thebibliography}{10}

\bibitem{de2020context}
E.~de~Matos, R.~T. Tiburski, C.~R. Moratelli, S.~Johann~Filho, L.~A. Amaral, G.~Ramachandran, B.~Krishnamachari, and F.~Hessel, ``Context information sharing for the internet of things: A survey,'' {\em Computer Networks}, vol.~166, p.~106988, 2020.

\bibitem{ramachandran2019towards}
G.~S. Ramachandran and B.~Krishnamachari, ``Towards a large scale iot through partnership, incentive, and services: A vision, architecture, and future directions,'' {\em Open Journal of Internet Of Things (OJIOT)}, vol.~5, no.~1, pp.~80--92, 2019.

\bibitem{tiburski2015importance}
R.~T. Tiburski, L.~A. Amaral, E.~De~Matos, and F.~Hessel, ``The importance of a standard securit y archit ecture for soa-based iot middleware,'' {\em IEEE Communications Magazine}, vol.~53, no.~12, pp.~20--26, 2015.

\bibitem{albouq2022survey}
S.~S. Albouq, A.~A. Abi~Sen, N.~Almashf, M.~Yamin, A.~Alshanqiti, and N.~M. Bahbouh, ``A survey of interoperability challenges and solutions for dealing with them in iot environment,'' {\em IEEE Access}, vol.~10, pp.~36416--36428, 2022.

\bibitem{casadei2019development}
R.~Casadei, G.~Fortino, D.~Pianini, W.~Russo, C.~Savaglio, and M.~Viroli, ``A development approach for collective opportunistic edge-of-things services,'' {\em Information Sciences}, vol.~498, pp.~154--169, 2019.

\bibitem{guide2021threatmodelling}
{Cyber Security Agency of Singapore}, {\em Guide to Conducting Cybersecurity Risk Assessment for Critical Information Infrastructure}.
\newblock Cyber Security Agency of Singapore, 2021.

\bibitem{Mitre}
``{MITRE ATT\&CK®}.'' \url{https://attack.mitre.org/}.
\newblock Accessed: 28 Sept 2023.

\bibitem{faieq2017c2iot}
S.~Faieq, R.~Saidi, H.~Elghazi, and M.~D. Rahmani, ``C2iot: A framework for cloud-based context-aware internet of things services for smart cities,'' {\em Procedia Computer Science}, vol.~110, pp.~151--158, 2017.

\bibitem{forkan2015bdcam}
A.~R.~M. Forkan, I.~Khalil, A.~Ibaida, and Z.~Tari, ``Bdcam: Big data for context-aware monitoring—a personalized knowledge discovery framework for assisted healthcare,'' {\em IEEE transactions on cloud computing}, vol.~5, no.~4, pp.~628--641, 2015.

\bibitem{huru2018bigclue}
D.~Huru, C.~Leordeanu, E.~Apostol, and V.~Cristea, ``Bigclue: Towards a generic iot cross-domain data processing platform,'' in {\em 2018 IEEE 14th International Conference on Intelligent Computer Communication and Processing (ICCP)}, pp.~427--434, IEEE, 2018.

\bibitem{fortino2018agent}
G.~Fortino, W.~Russo, C.~Savaglio, W.~Shen, and M.~Zhou, ``Agent-oriented cooperative smart objects: From iot system design to implementation,'' {\em IEEE Transactions on Systems, Man, and Cybernetics: Systems}, vol.~48, no.~11, pp.~1939--1956, 2018.

\bibitem{hassani2018context}
A.~Hassani, A.~Medvedev, P.~D. Haghighi, S.~Ling, M.~Indrawan-Santiago, A.~Zaslavsky, and P.~P. Jayaraman, ``Context-as-a-service platform: exchange and share context in an iot ecosystem,'' in {\em 2018 IEEE International Conference on Pervasive Computing and Communications Workshops (PerCom Workshops)}, pp.~385--390, IEEE, 2018.

\bibitem{liu2019scents}
C.~Liu, J.~Hua, and C.~Julien, ``Scents: Collaborative sensing in proximity iot networks,'' in {\em 2019 IEEE International Conference on Pervasive Computing and Communications Workshops (PerCom Workshops)}, pp.~189--195, IEEE, 2019.

\bibitem{zhang2020demand}
J.~Zhang, M.~Ma, W.~He, and P.~Wang, ``On-demand deployment for iot applications,'' {\em Journal of Systems Architecture}, vol.~111, p.~101794, 2020.

\bibitem{pereira2022platform}
J.~Pereira, T.~Batista, E.~Cavalcante, A.~Souza, F.~Lopes, and N.~Cacho, ``A platform for integrating heterogeneous data and developing smart city applications,'' {\em Future Generation Computer Systems}, vol.~128, pp.~552--566, 2022.

\bibitem{borges2023taming}
P.~V. Borges, C.~Taconet, S.~Chabridon, D.~Conan, E.~Cavalcante, and T.~Batista, ``Taming internet of things application development with the iotvar middleware,'' {\em ACM Transactions on Internet Technology}, vol.~23, no.~2, pp.~1--21, 2023.

\bibitem{papadakis2023comdex}
N.~Papadakis, G.~Bouloukakis, and K.~Magoutis, ``Comdex: A context-aware federated platform for iot-enhanced communities,'' in {\em DEBS 2023-17th ACM International Conference on Distributed and Event-Based Systems}, 2023.

\bibitem{cirillo2019standardFIWARE}
F.~Cirillo, G.~Solmaz, E.~L. Berz, M.~Bauer, B.~Cheng, and E.~Kovacs, ``A standard-based open source iot platform: Fiware,'' {\em IEEE Internet of Things Magazine}, vol.~2, no.~3, pp.~12--18, 2019.

\bibitem{perera2013context}
C.~Perera, A.~Zaslavsky, P.~Christen, and D.~Georgakopoulos, ``Context aware computing for the internet of things: A survey,'' {\em IEEE communications surveys \& tutorials}, vol.~16, no.~1, pp.~414--454, 2014.

\end{thebibliography}

\end{document}